\documentstyle[11pt]{article} 

\begin{document} 
\begin{flushright} 
June 13, 1996    
\end{flushright} 
  
\vskip 10pt
\begin{center} 
\Large\bf 
Relativistic Gravity Theory And Related Tests\\
With A Mercury Orbiter Mission

\vskip 26pt

\normalsize 

Slava G. Turyshev\footnote{ On leave from Bogolyubov Institute for 
Theoretical Microphysics, Moscow

\hskip 8pt State University, Moscow, 119899, Russia. 
Internet: sgt@zeus.jpl.nasa.gov}, 
John D. Anderson\footnote{
Internet: John.D.Anderson@jpl.nasa.gov} and Ronald W. Hellings\footnote{
Internet: rwh@graviton.jpl.nasa.gov}

\vskip 8pt

{\it Jet Propulsion Laboratory MS 301-230, \\
California Institute of Technology \\ 
 4800 Oak Grove Drive, 
Pasadena, CA 91109-8099, USA.}
\vskip 10pt
\end{center}

\begin{abstract}

Due to its relatively large eccentricity and  proximity
to the Sun, Mercury's orbital motion provides one of the best
solar-system tests of relativistic gravity.
We emphasize the number of feasible  
relativistic gravity tests that can be performed within 
the context of the parameterized weak field and slow motion 
 approximation    - a useful
framework for testing  modern gravitational theories in the solar system.
We discuss a new approximation method, which  
includes two Eddington parameters $(\gamma, \beta)$, proposed for construction
  of the relativistic equations of motion of  extended bodies.
Within the present accuracy of  radio  measurements, we discuss the generalized
Fermi-normal-like  proper reference frame  which is  defined in the 
immediate vicinity of the extended compact bodies.
Based on the Hermean-centric equations of motion of the spacecraft around 
the planet Mercury, we suggest a new test of the Strong Equivalence 
Principle. The corresponding experiment  
could be performed with the future {\it Mercury Orbiter} mission
scheduled by the European Space Agency ({\small ESA}) for launch between 2006 
and 2016.  We discuss other relativistic  effects including
the perihelion advance, redshift and geodetic precession of the orbiter's 
orbital plane   about Mercury.
 
\noindent PACS number(s): \hskip 2pt 95.10Ce, \hskip 2pt 04.20.Jb, 
\hskip 2pt 04.20.Me, \hskip 2pt 04.70.-s
\vskip 5mm
\end{abstract}

\section{Introduction}

It is well known that some  gravitational effects 
 depend  not only on the first or second derivatives 
of the gravitational potential, but 
also on the magnitude of this potential in the area where 
the gravitational experiments are performed. This is why, with 
the present level of experimental
techniques,   one should focus on  locations with an intensive 
gravitational environment  as the best place for conducting these studies.
The planet Mercury  plays  a   specific role in the history of 
modern gravitational physics. The successful  explanation 
of the perihelion advance of its orbit has become  one of the 
observational cornerstones for general relativity. 
Because of its proximity to the Sun, high eccentricity, and short 
orbital period, it  offers a very interesting opportunity 
for the study of relativistic gravity.
A spacecraft, placed in orbit 
about this planet,  can provide the additional 
data necessary for dynamic 
tests of the principles of  modern gravitational theories as well 
as the validation of the  approximation methods used for 
analysis of the gravitational environment.

In this paper we  analyze the   relativistic gravitational 
experiments for  the {\it Mercury Orbiter}  mission which  has been
included by the European Space Agency as a cornerstone mission under 
the Horizon 2000 Plus program. 
The motivation for the research described here is to determine what  
scientific information may be obtained during this mission, how accurate these 
measurements can be, and  what will be the significance of 
the  knowledge obtained. As it is known, there are three different 
types of measurements that are used in spacecraft navigation:
radiometric (range and Doppler), very-long baseline interferometry 
{\small (VLBI)} and optical \cite{Sta95}. In addition to navigation needs, 
the high precision Doppler, laser  and radio range  measurements of the velocity 
of and the distance to  celestial bodies  and  spacecraft are presently  
the best ways to collect  
important information about relativistic gravity within the solar system.
Combined with   the technique of ground- and space-based  
 {\small  VLBI}, these  methods provide us with a unique
opportunity to explore the physical phenomena in  our 
universe with  very high precision. Most remarkable is that the accuracy of the
 modern {\small VLBI} observations  is steadily increasing.  Thus  the delay 
residuals are presently of the order of 30-50 picoseconds (ps), which corresponds 
to an uncertainty in length of $\sim$ 1 cm.  Concerning the navigation of the 
interplanetary spacecraft, the short arcs of spacecraft range and Doppler 
measurements, reduced with Earth orientation information referred to the 
International Earth Rotation Service' {\small (IERS)} celestial system, 
lead to a position determination in the extragalactic reference frame
with an accuracy of order $\sim 20$ milliarcsecond (mas).
At the same time {\small VLBI} observations of the spacecraft with 
respect to extragalactic radio-sources enable a  direct measurement of one 
component of the spacecraft position in this extragalactic reference frame to 
an accuracy 
of about $\sim 5$ mas  (\cite{Bor82}, \cite{Fol94}).
As a result, the  use of such precise methods enables one to study the dynamics 
of celestial bodies and  spacecraft with an unprecedented high accuracy. 
These data    provide   the necessary 
foundation for research of many scientific problems, such as:
\begin{itemize}
\item[(i).] The construction of a dynamic  inertial 
 reference frame and a more precise definition of the orbital 
elements of the Sun, Earth, moon,  planets and their satellites 
 (\cite{Sta95}, \cite{Stand92}-\cite{Willi96}). 
\item[(ii).] The construction of  a kinematic 
inertial    reference frame, based on the observations of 
stars and quasars from  spaceborne astronomical observatories 
(\cite{Stand92}, \cite{Fuk91a}). 
\item[(iii).] The construction of a precise 
ephemeris for the motion of   bodies in the solar system
to  support   reliable navigation in the solar system  
(\cite{Sta95}, \cite{Stand95}). 
The construction of  precise radio-star catalogs for the spacecraft 
astroorientation and navigation  in outer space beyond the solar system.
\item[(iv).] The  comparison of   dynamic  and 
kinematic  inertial  reference frames,  based on the observations of 
 spacecraft on the  background  of quasars, pulsars and  
radio-stars, as well as the verification of the zero-points of 
the coordinates in the inertial  reference frame  (\cite{Fol94}, 
\cite{Jac93}, \cite{Fuk95}).
\item[(v).] The experimental tests of the weak field and slow motion 
 approximation ({\small WFSMA}) 
 of  modern theories of gravity (\cite{Dam83}, \cite{Wil93}).
\end{itemize}

These studies  will  enrich our knowledge  
about our universe, its cosmological evolution, and 
the behavior of  stellar systems in general. 
By presenting this list of  problems, we would like to emphasize the 
importance  of the {\it Mercury Orbiter} mission not only for the 
gravitational theory, but for many  of the  fundamental problems stated above. 
It should be noted  that, besides  offering the opportunity 
to study the gravitational environment in   Mercury's  vicinity,
this mission will also provide   important data  
about the gravitational field of the Sun, Mercury's magnetosphere and 
its interaction with the solar plasma (see more detailed analysis 
of these problems in \cite{AndTur96}),  as well as  
enable one to verify the foundations of many recently proposed theories of  
relativistic reference frames (\cite{Bru88a}-\cite{Tur96}).
This mission provides a good opportunity for
 testing  the principles of many modern
approximation formalisms   proposed  for describing  
gravitational wave generation mechanisms  (\cite{Dam83}, \cite{Bla95}). 
It should be noted  that  a physically justified definition of the proper  
reference frame of the extended bodies
({\it i.e.} a complete multipolar solution of the gravitational  {\small N}-body 
problem), in the considerably weak solar gravitational field, will  allow us to 
better understand the physical processes in a stronger gravitational field regime.  

This paper discusses the principles of a new approximation method developed for 
an astronomical relativistic  {\small N}-body problem and analyzes 
the  space gravitational experiments proposed for
 the future {\it Mercury Orbiter} mission.   The outline of the
paper is  as follows.  The next  Section contains a
brief introduction to the modern theoretical methods used to
describe the motion of a system of   {\small N}  weakly interacting  
self-gravitating extended bodies and to
analyze    gravitational experiments within the solar system. 
We discuss problems associated with the traditional   barycentric approach
of the {\small PPN} formalism. In Section {\small III} we discuss 
the principles of a new iterative method for constructing   the 
 solution to the gravitational 
{\small N}-body problem in the {\small WFSMA}. In particular we discuss the 
physical and mathematical properties of the relativistic astronomical reference 
frames. In Section {\small IV} this method is applied to derive gravitational 
field solutions and  coordinate transformations.   Our derivations include 
two Eddington parameters $(\gamma, \beta)$ which
allow  us to develop a new parametric theory of astronomical
reference frames. Within  the present accuracy of  radio  
measurements, we discuss the generalized
Fermi-normal-like  proper  reference frame,  which is  defined in the 
immediate vicinity of extended  bodies.
 Thus, we have obtained the interesting result  that 
although some terms in the Hermean-centric equations of motion of the spacecraft 
around the planet Mercury are zero for the case of general relativity, 
they may produce an observable effect  
in scalar-tensor theories. This  allows us to propose a new test of the   
  Strong Equivalence Principle ({\small SEP}), which is analyzed in   Section 
{\small V} of the paper.  Also in the  fifth Section we discuss  a number of 
relativistic gravitational experiments possible with the {\it Mercury Orbiter}. 
We  present there both a quantitative and 
a qualitative analysis of the measurable effects  such as Mercury's perihelion 
advance, the precession phenomena of the Hermean orbital plane, and the redshift 
experiment. In  Section {\small VI} we   present the  conclusions and 
recommendations for future gravitational experiments with 
the {\it Mercury Orbiter} mission. In order to make  access to the basic results 
of this paper easier, we present some calculations   in the Appendices.       

\section{Parametrized Post-Newtonian Gravity.}

Metric theories of gravity have a special position 
among all the other possible theoretical models.  
The reason  is that, independent  of the many different 
principles at their foundations, the gravitational field  
in these theories affects  matter directly through the metric 
tensor of the Riemann  space-time   $g_{mn}$, which is  
determined from the field equations of a particular theory of gravity. 
As a result, in contrast to Newtonian gravity, this tensor  
contains the properties of a particular gravitational theory as well as  
carrying the information about the gravitational field of the bodies. 
This property of the metric tensor enables one to analyze 
the motion of matter in one or another metric theory of
gravity  based only on the basic principles 
of modern theoretical physics.

Within the accuracy of modern experimental 
techniques, the   {\small WFSMA}  
provides a useful starting point for testing the
 predictions of different metric
theories of gravity in the solar system. 
Following Fock (\cite{Foc55}, \cite{Foc57}),  the perfect fluid 
is used most frequently   as the model of   matter distribution 
when describing   the   gravitational behavior of 
celestial bodies in this approximation.
The density of the corresponding energy-momentum  tensor $\widehat{T}^{mn}$ 
 is as follows\footnote{In this paper the notations are the same as in 
 \cite{Lan88}. In particular,  
the first three small latin letters  $a,b,c$   number the bodies
and run from 1 to $N$; the small 
latin letters $k,l, ...p$ run
from 0 to 3 and greek letters $\alpha, \beta, \gamma,...$ 
run from 1 to 3;  the comma denotes a standard partial 
derivative and semicolon denotes  a covariant derivative; repeat 
indices imply an Einstein rule of summation;
round brackets surrounding indices denote  
symmetrization and square brackets denote  
anti-symmetrization.}:
{}
$$\widehat{T}^{mn}=\sqrt{-g}\Big(\Big[\rho( 1 + \Pi) + 
p\Big]u^{m}  u^{n}  - p g^{mn} \Big), \eqno(1)$$ 

\noindent where $\rho$  is mass density of the ideal fluid in
 coordinates of the co-moving   reference frame, $u^k = {d z^k / d s}$ 
are the components of invariant four-velocity
of a fluid element, and $p(\rho)$ is the isentropic pressure 
connected with $\rho$ by an equation of state. The  quantity $\rho\Pi$ is the 
density of internal energy of an ideal fluid. The  definition of $\Pi$ is
 given by the equation based on the  first law of thermodynamics 
(\cite{Wil93}, \cite{Foc55}, \cite{Chandra65},   \cite{Bru72}):
{}
$$u^n\Big(\Pi_{;n}  + p\Big({1\over {\widehat \rho}}\Big)_{;n}\Big) = 0,\eqno(2)$$

\noindent where ${\widehat \rho}=\sqrt{-g}\rho u^0$ is the conserved mass density. 
Given the energy-momentum tensor, one may proceed   to find the solutions of
the gravitational field equations for a particular relativistic theory of gravity.  
The solution for an astronomical   {\small N}-body problem is the one of most  
practical interest.  In the following  Sections we will discuss the  properties 
of the  solution of an isolated one-body problem as well as the features  of 
construction of the general solution for the  {\small N}-body problem in both  
barycentric and planeto-centric reference frames. 

\subsection{An Isolated One-Body Problem.}

The solution for the isolated one-body problem in  the {\small WFSMA} 
may be obtained from the linearized gravitational field equations
of a particular theory under study. 
A perturbative gravitational field $h_{(0)}^{mn}$, in this case,  is 
characterized by 
the deviation of the density of the 
general Riemmanian metric tensor $\sqrt{-g}g^{mn}$ from the background 
space-time $\eta_{mn}$, which is considered to be a zero-th
order\footnote{For most of the non-radiative problems  in  solar system 
dynamics, this tensor usually is  taken to be a  flat Minkowski  metric
(\cite{Dam83}, \cite{Wil93}).}  approximation 
for a series of   successive iterations:
$\sqrt{-g}g^{mn}-\sqrt{-\eta}\eta_{mn}=h_{(0)}^{mn}$, or equivalently 
{}
$$g_{mn}=\eta_{mn}+h^{(0)}_{mn}. \eqno(3a) $$ 

In order to accumulate the features of many modern metric theories of 
gravity in one  theoretical scheme, as well as 
to create a  versatile mechanism to plan   gravitational 
experiments and to analyze the 
data obtained,  Nordtvedt and Will have proposed 
the parameterized post-Newtonian 
({\small PPN}) formalism (\cite{Nor68a}-\cite{Wil72}).  
This formalism allows one to describe  the motion of 
celestial bodies  for a wide class of metric 
theories of gravity within a
common framework.  The gravitational field in the  {\small PPN}  formalism is
presumed to be generated by some isolated distribution of matter 
which is taken to be an ideal fluid Eq.(1). This field is represented by a sum of  
gravitational potentials with   arbitrary coefficients:  the 
  {\small PPN}  parameters. The two-parameter  form of this  
tensor in four dimensions   may be written as follows:

$$h^{(0)}_{00}= -2U+2(\beta-\tau) U^2+2\Psi+ 2\tau(\Phi_2-\Phi_w)+
(1-2\nu) \chi,_{00}+{\cal O}(c^{-6}), \eqno(3b)$$
{}
$$h^{(0)}_{0\alpha}=(2\gamma+2-\nu-\tau)V_\alpha+ 
(\nu+\tau)W_\alpha+{\cal O}(c^{-5}),\eqno(3c)$$
{}
$$h^{(0)}_{\alpha\beta}=2\eta_{\alpha\beta} (\gamma-\tau) U -
2\tau U_{\alpha\beta}+  {\cal O}(c^{-4}),\eqno(3d)$$

\noindent where $\eta_{mn}$ is the Minkowski 
metric\footnote{The following  metric convention $(+---)$ is used 
throughout  as is the geometrical units $\hbar=c=G=1$.} 
and the generalized gravitational potentials are given
in  Appendix A.

Besides the  two Eddington  parameters 
$(\gamma,\beta)$,  
Eq.(3) contains   two other parameters  $\nu$ and $\tau$.  
The  parameter $\nu$ reflects the specific
choice of the gauge conditions. For the standard   
 {\small PPN}  gauge it is  given as  $\nu= {1\over 2}$, but for harmonic gauge 
conditions one should choose $\nu=0$. The parameter $\tau$ 
describes  a possible pre-existing anisotropy of   space-time and  
corresponds to different  spatial coordinates, which may be  chosen for 
modelling the experimental situation. For example, the case $\tau=0$ 
corresponds to harmonic coordinates, while $\tau=1$ corresponds to the 
standard (Schwarzschild) 
coordinates. A particular metric theory of gravity in this framework
with a  specific coordinate  
gauge $(\nu,\tau)$ may then be characterized by  means of  
two   {\small PPN}  parameters $(\gamma, \beta)$, which are uniquely prescribed 
for each particular theory under study.  
In the  standard  {\small PPN}  gauge 
(i.e. in the case when $\nu={1\over 2}, \tau=0$) these  parameters have clear 
physical meaning.
The parameter $\gamma$ represents the measure of the curvature of the space-time 
created by the unit rest mass; the parameter $\beta$ is the measure of the 
non-linearity of the law of superposition of the gravitational fields in the
theory of gravity (or the measure of the metricity). 
Note that general relativity, when analyzed in  standard 
 {\small PPN}  gauge, gives: $\gamma=\beta=1$,
whereas, for  the Brans-Dicke theory, one has 
$\beta=1, \gamma= {1+\omega\over 2+\omega}$, where $\omega$ 
is an unspecified dimensionless parameter of the theory. 
 
The properties of an isolated  one-body solution  are well-known. 
It has been shown (\cite{Wil93}, \cite{Lee74},  \cite{Ni78})
that for an  isolated distribution of matter in  {\small WFSMA}   
 there exist  a set of inertial reference frames  and ten integrals
of motion  corresponding to ten conservation laws. Therefore, 
it is possible to consistently define  the multipole moments characterizing the
body under study. For  practical purposes one   chooses the inertial 
reference frame located in the center of mass of an 
isolated distribution of matter. By performing a power expansion 
of the potentials in terms of
spherical harmonics, one may obtain  the  
post-Newtonian  set of  `canonical' parameters (such as  unperturbed irreducible 
mass $I^{\{L\}}_{a(0)}$ and current $S^{\{L\}}_{a(0)}$ 
multipole moments\footnote{To enable one to deal conveniently with sequences of 
many spatial indices, we shall use an abbreviated notation for `multi-indices' 
where an upper-case letter in curly brackets denotes a multi-index, 
while the corresponding lower-case  letter denotes its number
of indices, for example: {\small $\{P\}:=\mu_1\mu_2 \ldots \mu_p$, 
$S_{\{J\}}:= S_{\mu_1\mu_2\ldots \mu_j}$}. 
The explicit expression for the symmetric and trace-free {\small (STF)} part 
of the tensor $T_{\{P\}}$ is given in 
(\cite{Tho80}-\cite{Bla89}). }),
generated by the inertially moving extended 
self-gravitating body  {\small (A)} under consideration:
{}
$$I^{\{L\}}_{a(0)}=\Big[\int_a d^3z'_a {\hat t}^{00}_a(z'^p_A)  
z'^{\{L\}}_a\Big]^{\it STF}\hskip -15pt, \qquad \hskip 10pt
 S^{\{L\}}_{a(0)} =\Big[\epsilon^{\mu_1}_{\hskip 4pt\beta\sigma} \int_ad^3z'_a 
z'^\beta_a{\hat t}^{0\sigma}_a(z'^p_a) 
z'^{\mu_2}_a...z'^{\mu_l}_a\Big]^{\it STF}\hskip -15pt, \eqno(4a)$$  
\noindent where ${\hat t}^{mn}_a$ are the components of the symmetric density
of the energy-momentum tensor of matter and gravitational field taken jointly.
As a result, the corresponding gravitational field $h_{(0)}^{mn}$ 
may be uniquely represented  in the external domain  as a 
functional depending on the set of these moments. 
Schematically this may be expressed as:
{} 
$$h_{(0)a}^{mn}={\cal F}^{mn}\big[I^{\{L\}}_{a(0)}, S^{\{L\}}_{a(0)}\big],
\eqno(4b)$$ \noindent  where the functional
dependence, in general, includes a non-local time dependence on the `past' 
history\footnote{Gravitational radiation problems are not within   
the scope of the present paper, and hence the  set of multipole moments  
Eq.(4)  are
used for both tensor and scalar-tensor theories.} of the 
moments \cite{Bla95}. However, by assuming   
that the  internal processes in the body are adiabatic, one may neglect this 
non-local evolution. As a result,   an external observer may 
uniquely establish the gravitational field of this body through  
determination of these multipole moments, for example,  by studying the geodesic 
motion of the test particles in orbit around this distribution of matter
 \cite{Mis73}. 
 
 It has been shown (\cite{Foc55},  \cite{Lee74}, \cite{Ni78})
 that for an  isolated distribution of matter in  {\small WFSMA}   
it is possible to consistently define the lowest conserved 
multipole moments, such as the rest mass of  the  
body  $m_a$, it's center of mass $z^{\alpha}_{a_0}$, 
momentum $p^{\alpha}_a$ and angular momentum 
$S^{\alpha\beta}_a$. Thus the definitions for the mass 
$m_a$ and coordinates of the center of mass of the
 body $z^{\alpha}_{a_0}$  in any inertial 
reference frame  are given by the formulae (for more 
detailed analysis see (\cite{Wil93}, \cite{Dam83}, \cite{Dam87}
and references therein):
{}
$$m_a = \int_a d^3 z'^{\nu }_a  \hskip 1mm{\hat t}^{00}(z'^p),
\qquad z^{\alpha}_{a_0}(t) =  {1\over m_a }  \int_a d^3 z'^{\nu }_a
\hskip 1mm        z'^{\alpha}_a  {\hat t}^{00}(z'^p). \eqno(5a)$$ 
\noindent where the energy density ${\hat t}^{00}(z'^p)$ of 
the matter and the gravitational field is given by:
$$  {\hat t}^{00}(z'^p)= \widehat{\rho}_{a}\Big( 1 +\Pi  + {1\over2}  U -   
{1\over2}{\underline v}_\mu {\underline v}^\mu
+{\cal O}(c^{-4}) \Big), \eqno(5b)$$  
\noindent with $\widehat{\rho}_{a}$ being the conserved mass density  defined by  
$\widehat{\rho}_{a}={\rho_a}  \sqrt{-g} u^0$ and   ${\underline v}^\mu$   the 
velocity of the intrinsic motion of matter.
In particular,  the center of  mass 
$z^{\alpha}_{a_0}$  moves in 
space with a constant velocity along a straight line:
 $z^\alpha_{a_0}(t) = p^\alpha_a \cdot t + k^\alpha_a$,
where the constants $p^\alpha_a = d z^{\alpha}_{a_0} /d t$ 
and $k^\alpha_a$ are the body's momentum and center of 
inertia, respectively. One may choose from the set of 
inertial reference frames the barycentric inertial one. In this frame the 
functions 
$z^\alpha_{a_0}$ must equal zero for any moment of time. 
This condition may be satisfied by applying to the metric
Eq.(3) the post-Galilean transformations  \cite{Chandra67}
 where  the constant velocity  and displacement of origin  
should be selected in  such a way that $p^\alpha_a$ and $k^\alpha_a$ 
equal zero (for details see  \cite{Wil93}, \cite{Kop88}). 

\subsection{An Astronomical $N$-Body System.}

By putting some restrictions on the shape and internal structure of the 
bodies, one can generalize   the results presented above to the case of 
an isolated  astronomical  {\small N}-body system. 
Indeed, the  assumption that the bodies posses only the lowest multipole 
moments  (such as mass $m_a$, intrinsic spin moment 
$S^{\alpha\beta}_a$ and the quadrupole moment $I^{\alpha\beta}_a$)  
 considerably simplifies the problem. 
The general solution   with such an assumption is well-known 
(\cite{Foc55}, see \cite{Dam86} and  references therein).
The main properties  of the solution Eq.(3)  applied to such a case 
are well  established and widely in use in modern 
ephemerides astronomy.
In order to analyze the  motion of bodies in the solar system 
barycentric reference frame,  one may obtain the restricted Lagrangian 
function $L_N$ describing the motion of  {\small N}  self-gravitating 
bodies (\cite{Wil93}, \cite{Tur90}). 
Well within the accuracy necessary for our future discussion of the 
gravitational experiments on the {\it Mercury Orbiter} mission, this  
simplified function may be presented  in the following form:
{} 
$$L_N=\sum_a^N{m_a\over 2}{v_a}_\mu{v_a}^\mu 
\Big(1-{1\over 4}{v_a}_\mu{v_a}^\mu\Big)- 
\sum_a^N\sum_{b\not=a}^N{m_a m_b\over r_{ab}}\Bigg({1\over2}
+(3+\gamma-4\beta)E_a-$$
{}
$$-(\gamma-\tau+{1\over 2}){v_a}_\mu {v_a}^\mu +
(\gamma-\tau+{3\over 4}){v_a}_\mu {v_b}^\mu -({1\over 4}+\tau) 
{n_{ab}}_\lambda {n_{ab}}_\mu 
{v_a}^\lambda {v_b}^\mu+\tau ({n_{ab}}_\mu  {v_a}^\mu)^2+$$
{}
$$+{{n_{ab}}_\lambda\over r_{ab}} \Big[(\gamma+{1\over 2}){v_a}_\mu-
 (\gamma+1){v_b}_\mu\Big] S_a^{\mu\lambda} + {n_{ab}}_\lambda 
{n_{ab}}_\mu {I_a^{\lambda\mu}\over r_{ab}^2}\Bigg)+
(\beta-\tau-{1\over 2}) \sum_a^N m_a\Big(
\sum_{b\not=a}^N {m_b\over r_{ab}}\Big)^2- $$
{}
$$-\tau\sum_a^N\sum_{b\not=a}^N \sum_{c\not=a,b}^N
m_a m_bm_c\Big[ {{n_{ab}}_\lambda\over 2r_{ab}^2}
(n_{bc}^\lambda+n_{ca}^\lambda)-{1\over r_{ab}r_{ac}}\Big]+
\sum_a^N m_a {\cal O}(c^{-6}), \eqno(6) $$ 

\noindent  where $m_a$ is the isolated rest mass of a body $(a)$,
the vector $r_a^\alpha$ is the barycentric radius-vector of this 
body, the vector $r_{ab}^\alpha=r_b^\alpha -r_a^\alpha$ is the 
vector directed from body $(a)$ to body  $(b)$, and
the vector $n^\alpha_{ab}=r^\alpha_{ab}/r_{ab}$ is the usual notation 
for the unit vector along 
this direction. It should be noted that Eq.(6) 
does not depend on the parameter $\nu$, which confirms that 
this parameter is a gauge parameter only. 
The tensor $I_a^{\mu\nu}$ is  the  quadrupole moment of 
body $(a)$  defined as:
{}
$$I_a^{\mu\nu}={1\over 2m_a}\int_a d^3z'^\nu_a\widehat{\rho}_{a}(z'^p_a) 
\Big(3z'^\mu_a z'^\nu_a-\eta^{\mu\nu}z'_{a\beta} z'^\beta_a \Big).  \eqno(7)$$

\noindent  The tensor $S_a^{\mu\nu}$ 
is the body's intrinsic   spin moment  which is given as:
{}
$$S_a^{\mu\nu}={1\over m_a}\int_a  d^3z'^\nu_a \widehat{\rho}_a(z'^p_a)
\big[{\underline v}_a^\mu z'^\nu_a-{\underline v}_a^\nu z'^\mu_a\big], \eqno(8)$$

\noindent where ${\underline v}_a^\mu$ is the 
velocity of the intrinsic motion of matter in  body $(a)$. Finally, the 
quantity $E_a$ 
is the body's gravitational binding energy:
{}
$$E_a  = {1\over 2m_a}{\int\int}_a
d^3z'^\nu_a d^3z''^\nu_a  {\widehat{\rho}_a(z'^p_a) 
\widehat{\rho}_a(z''^p_a)\over|z'^\nu_a-z''^\nu_a|}. \eqno(9)$$

\noindent The corresponding  equations of motion of the planet
in the solar system barycentric reference frame  are 
given in   Appendix B. 
It should be noted here that  in the present numerical  algorithms  
for  celestial mechanics problems (\cite{Stand92},  \cite{Wil93}, 
\cite{Moy71}-\cite{Bru91}) the bodies in the solar system 
are assumed to possess  the lowest post-Newtonian multipole moments only. 
The corresponding  barycentric inertial reference frame   has been 
adopted for the fundamental
planetary and lunar ephemeris (\cite{Stand92}, \cite{New83}). Moreover, the 
coordinate time of 
the solar barycentric (harmonic) reference frame 
is the {\small TDB} time scale, which has been  adopted by  Fukushima 
\cite{Fuk95} and Standish \cite{Sta92}.

However,   it turns out that   generalization of the 
 results  obtained for the one-body problem, to
 the problem of motion of an arbitrary {\small N}-body system is not   quite 
straightforward.  
Thus, we point out that taking into account the presence of any 
non-vanishing internal multipole moments of an extended body,
significantly changes  its equations of motion due to  
coupling of the intrinsic multipole moments of the body to the 
surrounding gravitational field. 
For example,  for a   neutral monopole test particle, the external 
gravitational field  completely defines the  fiducial geodesic 
world line  which this test body follows (\cite{Wil93}, \cite{Foc55}). 
On the other hand, the equations of motion for spinning  bodies differ from 
these by additional terms due to the coupling of the  body's spin  
to the external gravitational field (\cite{Pap48}-\cite{Bar75}). 
As a result, one needs to present a post-Newtonian definition for the proper
intrinsic multipole moments of the bodies  in order to describe   their 
interaction with the surrounding gravitational field and to   obtain the  
corrections to the laws of motion and to the precession of extended bodies in 
this system (\cite{Tho88}-\cite{KliVoi93}). The fully relativistic 
definition of these   
 moments may be  given in the proper quasi-inertial reference frame only. 
Such a definition   replaces  Eq.(4) which was given in the rest-frame of the 
one-body problem\footnote{Note that, due to the breaking of the symmetry 
of the total Riemann space-time by realizing the $3+1$ split  
\cite{Tho88}, these moments will not form  tensor quantities  
with respect to general  four-dimensional coordinate transformations.
Instead, these quantities will behave as  tensors under the sub-group of this 
total group of motion only, namely: the three-dimensional  rotation. 
This is similar to the situation in classical electrodynamics, where   electric 
${\vec E}$ and magnetic ${\vec H}$ fields are not   true vectors, 
but  rather   components of the $4\times 4$ tensor 
of the electro-magnetic field $F_{mn}=({\vec E}\otimes {\vec H})$ 
\cite{Lan88}.}. In presenting these transformations one should also 
take into account that, due to the non-linear character of the 
gravitational interaction, these moments are expected to interact  with   external
 gravity,  changing the state of motion of the body itself. 
 Fock \cite{Foc55} was the first to notice  that in order to find  the 
solution of the {\it global} problem (the motion of the {\small N}-body system 
as a whole),  the solution for the {\it local} gravitational problem (in the 
body's vicinity) is required. 
In addition, one must establish their correspondence by presenting 
the coordinate transformation by which the 
physical characteristics of motion and rotation are transformed 
from the coordinates of one reference frame  to another. 
Thus,   one must find   the solutions to the three following problems 
(\cite{Dam91}, \cite{Dam87}):

\begin{itemize}
\item[(1).]   \underline{  The {\it global} problem:} 
\begin{enumerate}
\item[(i).] We must  construct  the asymptotically  inertial reference frame.
\item[(ii).] We must  find  the barycentric inertial reference frame for the 
system under study. This is primarily a problem of describing  the {\it global} 
translational motion of the bodies constituting the {\small N}-extended-body  
system ({\it i.e.} finding the geodesic structure of the space-time occupied by 
the whole system).  
\end{enumerate}
\item[(2).]    \underline{ The {\it local} problem:}  
\begin{enumerate}
 \item[(i).]  We must   establish   the properties of 
the gravitational environment in the proximity of each body in the system 
 ({\it i.e.} finding the geodesic structure of the {\it local} region of the 
space-time in the body's gravitational domain). 
\item[(ii).] We must construct the  {\it local} effective rest frame  of each 
body.
\item[(iii).]  We must study     the internal motion 
of matter inside the bodies as well as   establish    their explicit 
multipolar structure and  rotational motion.

\end{enumerate}
 \item[(3).]  \underline{The theory of the reference frames:} 
\begin{enumerate}
 \item[(i).]
We must find a way to describe the  mutual physical cross-interpre\-tation 
 of the results obtained for the above two  problems ({\it i.e.}  the mapping 
 of the  space-time).  
\end{enumerate}
\end{itemize}
 
\noindent 
Because the   solutions to  the first two problems will not be complete without 
presenting the rules of  transforming between 
the {\it global} and the {\it local}    reference frames chosen 
for such an analysis, the theory of the astronomical 
reference frames become inseparable from the 
problem of determination of the motion of the celestial bodies. 
>From the other side, if one attempts to describe the {\it global} dynamics of the 
system of {\small N} arbitrarily shaped extended bodies, one 
will discover that even in  {\small WFSMA}  
this solution will not be possible without an
appropriate description of the gravitational environment 
in the  immediate vicinity of the bodies (\cite{Dam91}, \cite{Kop88}).  

\section{Gravitational N-body Problem in the WFSMA.}

In this Section we will discuss the principles of  a new iterative method for 
generating the solutions to an arbitrary   {\small N}-body gravitational 
problem in the {\small WFSMA} of  metric theories of gravity. 
In these theories  one may choose 
any coordinates to  describe  the gravitational environment around the 
body under question. However non-optimal choice of these coordinates 
may cause unreasonable complications in the  physical interpretations 
of the data obtained  (see the related  discussion in \cite{Kop88}, \cite{Sof91}).  
Recently several different attempts have been made to  
improve the present solution to the  {\small N}-body problem 
in  {\small WFSMA} 
(see, for example, \cite{Bru88a}-\cite{Dam91},  \cite{Kop88}, \cite{Ash84}-\cite{Kli93}). 
Note  that although these methods represent a significant improvement 
in our understanding of the general problem, the present situation requires 
a more physically adequate   approach   to account for the difficulties 
in describing the relativistic motion of a system of extended bodies.

In the present paper we use an  
approach which has being developed to deal with the complications introduced 
by using a  
standard  {\small PPN}  formalism  \cite{Tur96}.  
This new method is based on the Fock-Chandrasekhar approach for dealing 
with  extended and arbitrarily shaped celestial bodies  and 
is parameterized with the two Eddington parameters $(\gamma, \beta)$.
  This formalism is  based upon the construction of   proper 
 reference frames  in the vicinities of  each body in the system. 
Such  frames are defined in the gravitational domain 
occupied by a particular body $(b)$.  
One may expect that,  in the immediate vicinity of this body, its proper 
gravitational field will dominate, while  the existence of 
external gravity will   manifest   itself 
in the form of the tidal interaction only. Therefore, in  the case of 
the {\small WFSMA} and in extreme proximity to the 
body under study, this proper reference frame  should resemble  an 
inertial frame and the solution Eq.(3) for an isolated one-body problem 
$h_{mn}^{(0)b}$   
should adequately represent the physical situation. 
However, if one decides to perform a physical  experiment at some distance from 
the world-tube
of the body, one should consider the   existence of external gravity as well. 
This is true  because  external gravity  plays a more  significant role at 
large  distances from the body.

\subsection{Definition of a Physically Adequate Proper Reference Frame.}

In order to construct a general solution for the {\small N}-body problem 
in a metric theory of gravity, let us  make a few assumptions. First of all, 
we shall assume that the  solution of the gravitational field equations 
$h_{mn}^{(0)}$ 
for an isolated unperturbed distribution of matter  is known  and it 
is given by  the equations  (3). 
With this assumption, we may construct the total solution of the 
{\it global} problem $g_{mn}$ in an arbitrary  reference frame as a formal 
tensorial 
sum of the background space-time metric $\eta_{mn}$, the  
 unperturbed solutions $h_{mn}^{(0)b}$ plus the gravitational interaction term
${\cal H}_{mn}$. 
Thus,  in the coordinates $x^p\equiv (x^0, x^\nu)$ 
of the barycentric inertial reference frame, 
one  may search for the desired total solution in the following form
\cite{Tur94}:
{}
$$g_{mn}(x^p) =\eta_{mn}(x^p) + h_{mn}(x^p) = $$
$$= \eta_{mn}(x^p) + 
\sum_{b=1}^N {{\partial y^k_b}\over{\partial x^m}}
{{\partial y^l_b}\over{\partial x^n}}
\hskip 0.5mm h^{(0)b}_{kl}(y^q_b (x^p)) + 
{\cal H}_{mn}(x^p),  \eqno(10)$$

\noindent where  the coordinate transformation 
functions $y^q_b = y^q_b (x^p)$ are yet  to be determined.
The interaction term ${\cal H}_{mn}$ will be discussed below.

If the bodies in the system are compact and well separated, 
then, we may take into account that
the mutual gravitational interaction  between the bodies affects their  
distribution of matter  through the metric tensor only.
 Therefore, without any loss of  accuracy, we obtain the  
total  energy-momentum tensor 
of the matter distribution in the  system in the following form:     
{}
$$T_{mn}(x^s) =  \sum_{b=1}^N {{\partial y^k_b}
\over{\partial x^m}}{{\partial y^l_b}\over{\partial x^n}}
T^B_{kl}(y_B(x^s)), \eqno(11)$$
 
\noindent where $T^b_{kl}$ is the  energy-momentum 
tensor\footnote{As a partial result of the representation Eq.(11) one can see 
that   the Newtonian mass density $\rho_b$ 
of a particular body $(b)	$  is defined in a sense of a three-dimensional 
Dirac delta-function. Thus in the body's proper    compact-support 
volume one will have: $\rho_b=m_b\delta(y^\nu_b)$, so that 
$$\int_{V_a}d^3y'_a \rho _b(y'^p_a) = m_b \delta_{ab},$$
\noindent where $\delta_{ab}$ is the three-dimensional Kronekker symbol 
($\delta^a_b=\delta_{ab}$;  $\delta_{ab}=1$ for $a=b$ and $=0,$ for $a\not=b$).
Then in any proper reference frame  the total density ${\overline\rho}$ of the 
whole  {\small N}-body system 
will be given by the expression ${\overline\rho}(y^p_a)=\sum^N_b \rho _b(y^p_a)$.
This representation allows one to distinguish between the microscopic and 
macroscopic (or integral)
descriptions of the physical processes and, hence, provides correct  
 relativistic treatment of the problem of motion of  an astronomical  
{\small N}-body system.}  of a body $(b)$ as seen by a co-moving observer. 

One may establish the properties of the solution  Eq.(10) with respect to  an 
arbitrary coordinate  transformation  simply by applying the basic rules 
of tensorial coordinate transformations.
In particular,  in   the coordinates $y^p_a\equiv   (y^0_a, y^\nu_a)$ of an
arbitrary   proper reference frame   this tensor  
 will take the following form:
{}
$$g^a_{mn}(y^p_a) = {{\partial x^k}\over{\partial y^m_a}} 
{{\partial x^l}\over{\partial y^n_a}} \hskip 1mm 
g_{kl}(x^s(y^p_a)) =    {{\partial x^k}\over{\partial y^m_a}}{{\partial x^l}
\over{\partial y^n_a}}\hskip1mm \eta_{kl}(x^s(y^p_a))  +$$
{}
$$+ h^{(0)a}_{mn}(y^p_a) + \sum_{b\not=a} {{\partial y^k_b}
\over{\partial y^m_a}}{{\partial y^l_b}\over{\partial y^n_a}}
\hskip 0.5mm h^{(0)b}_{kl}(y^{s}_b (y^p_a))+{{\partial x^k}
\over{\partial y^m_a}}{{\partial x^l}
\over{\partial y^n_a}}\hskip1mm{\cal H}_{kl}(x^s(y^p_a)). \eqno(12)$$

\noindent The expression  for  $T_{mn}(y^p_a)$ 
could be  obtained analogously from that given by Eq.(11).  
To complete the formulation of the perturbative
 scheme we need to introduce the procedure  for  constructing
 the solutions for the various  unknown 
functions entering Eqs.(10)-(12), 
including the four functions of the coordinate 
transformations  $y^m_b = y^m_b(y^p_a)$ and the interaction term ${\cal H}_{mn}$.

 For the four functions of the coordinate transformations $y^m_b = y^m_b(y^p_a)$
we use the general post-Newtonian coordinate transformations  \cite{Tur96}
which  connect  the coordinates $(x^p)$ of the  
barycentric inertial reference frame    to those  
$(y^p_a)$ of a proper  quasi-inertial   
reference frame  of an arbitrary body $(a)$. The general 
form of these  relations may be given as follows:  
{}
$$x^0 = y^0_a+ c^{-2}{\cal K}_a(y^0_a, y^\epsilon_a) + 
c^{-4} {\cal L}_a(y^0_a, y^\epsilon_a) + {\cal O}(c^{-6}),
\eqno(13a)  $$
$$x^\alpha =  y^\alpha_a + y^\alpha_{a_0} (y^0_a) + 
c^{-2}{\cal Q}^\alpha_a(y^0_a,y^\epsilon_a) + {\cal O}(c^{-4}), 
\eqno(13b)$$
\noindent where the barycentric radius-vector $r^\alpha_a$ 
of the body $(a)$ in the coordinates of the proper 
reference frame  is decomposed into 
Newtonian and post-Newtonian parts and is given  
as  follows: 
{}
$$\Big<r^\alpha_{a_0}(y^p_a)\Big>_a = y^\alpha_{a_0} (y^0_a)+ 
{1\over m_a c^2}\int_a d^3y'^\nu_a
{\hat t}^{00}(y'^p_a) {\cal Q}^\alpha_a(y^0_a,y^\epsilon_a)+
{\cal O}(c^{-6}),\eqno(14a)$$

\noindent with 

$$m_a =\int_a d^3y'^\nu_a {\hat t}^{00} (y'^p_a) +{\cal O}(c^{-6}),
\eqno(14b)$$
\noindent where ${\hat t}^{00}(y'^p_a)$ is the component of the 
conserved density of the energy-momentum tensor of matter, inertia and  
gravitational field taken jointly. The notation  
$\big<..\big>_a$ applied to   any function $f(y^p_a)$ 
means an averaging over body  $(a)$'s   volume   as follows: 
{}
$$ \Big<f(y^p_a)\Big>_a \equiv {\hat f}(y^0_a)={1\over m_a}
\int_a d^3y'^\nu_a\hskip 2pt{\hat t}^{00}_a(y'^p_a)
f(y^0_a,y'^\nu_a). \eqno(15)$$   
 
In order to complete the formulation of the perturbative
scheme, we need  to introduce the  procedure for  constructing
the solutions for  the interaction term ${\cal H}_{mn}$  and  for the 
post-Newtonian transformation functions ${\cal K}_a, {\cal Q}^\alpha_a $ and 
${\cal L}_a$ which are still  unspecified. 
The way to construct the solution for the interaction term 
${\cal H}_{mn}$ is quite straightforward; it is sufficient to 
require that the metric tensor in the form of Eq.(10) or Eq.(12) 
will be the explicit solution   of the gravitational field equations in the 
corresponding reference frame. These solutions  
are assumed to satisfy the covariant 
harmonical de Donder
gauge, which,  for an arbitrary reference frame,   may be written  
as follows: 
{}
$${\cal D}^b_n\Big({\sqrt{-g}}_bg^{mn}_b(y^p_b)\Big) = 0, \eqno(16a)$$ 

\noindent 
where ${\cal D}^b_n$ is the covariant derivative with respect to the metric 
$\eta^b_{mn}(y^p_b)$ of the inertial Riemann-flat 
$\big(R^k_{nml}(\eta^b_{st}(y^p_b))=0\big)$ space-time in these 
coordinates\footnote{In Cartesian coordinates of the inertial Galilean 
reference frame   $(x^p)$ the flat metric $\eta_{mn}$ can be chosen as 
$\eta^{(0)}_{mn} = \hbox{diag}(1,-1,-1,-1)$, so that the Christoffel
 symbols $\Gamma^{k(0)}_{mn}  = 0$ all vanish  and  
conditions Eq.$(16a)$  take the more familiar form 
of the harmonic conditions 
{}
$$\partial_{n}\big({\sqrt{-g}}g^{mn}\big) = 0,$$
\noindent which  are equivalent to setting $\nu=\tau=0$ in the Eq.(3).}. 
For most of the practically interesting problems in the {\small WFSMA} 
in quasi-Cartesian coordinates 
this  metric may be  represented as the sum of two tensors: the Minkowski metric 
$\eta^{(0)}_{mn}$ and the field of  inertia $\phi_{mn}$:
{}
$$\eta^b_{mn}(y^p_b) = {{\partial x^k}\over{\partial y^m_b}}
{{\partial x^l}\over{\partial y^n_b}}\eta_{kl}(x^s(y^p_b))=
\eta^{(0)}_{mn}+\phi^b_{mn}(y^p_b).\eqno(16b)$$ 
\noindent Note that the term $\phi_{mn}$ appears to be 
parameterized by the coordinate transformation functions 
${\cal K}_a, {\cal L}_a$ and ${\cal Q}^\alpha_a$ defined in Eq.(13); 
thus we have $\phi_{mn}(y^p_a)=\phi_{mn}\big[{\cal K}_a,{\cal L}_a,
{\cal Q}^\alpha_a\big]$,
a formulation which will be referred to as the ${\cal KLQ}$ parameterization 
in the {\small WFSMA}.

The search for the general solution for ${\cal H}_{mn}(x^p)$ 
is performed in a barycentric inertial reference frame   $(x^p)$, 
which is singled out by the Fock-Sommerfeld's boundary conditions 
imposed on $h_{mn}$ and $\partial_k h_{mn}$ (given by Eq.(10)):
{}
$$\lim_{{r \rightarrow \infty} } 
\Big( h_{mn}(x^p); \hskip 2mm r\Big[{\partial \over \partial x^0}
 h_{mn}(x^p) + 
{\partial \over \partial r}  h_{mn}(x^p)\Big]\Big) 
\hskip 2mm \rightarrow \hskip 2mm 0, $$
$$ t + {r\over c} = {\it const},  \eqno(17a)$$

\noindent where $r^2 =  - \eta_{\mu\nu} \hskip 1mm x^\mu x^\nu$.
These conditions   define the asymptotically Minkowskian space-time
 in a weak sense, consistent with  the absence of any flux 
of gravitational radiation falling on the system from the external
 universe \cite{Dam83}. 
Moreover, one assumes that  there exists such a quantity 
$h_{mn}^{\it max} = {\it const}$ (for the solar system this constant
is of order $10^{-5}$) for which the condition  
{}
$$ h_{mn}(x^p) \hskip 1mm < \hskip 1mm h_{\it mn}^{max},  \eqno(17b)$$

\noindent 
should be satisfied for each point $(\vec{x})$  inside 
the system.  
Note  that any distribution of matter is considered 
isolated, if  conditions Eq.(17) are
fulfilled in any inertial reference frame.

The  advantage  of  using   the imposed conditions Eq.(17) 
is that it gives us the opportunity to determine   the interaction term 
${\cal H}_{mn}(x^p)$ in a unique way. It should be stressed that the 
 corresponding solution $g_{mn}(x^p)$ in the barycentric inertial reference frame  
resembles  the form of the solution for an isolated one-body problem Eq.(3).
The only change that should be made is to take into account the 
number of the bodies in the system: $\rho\rightarrow \sum_b\rho_b$, where $\rho_b$
is the compact-support mass density of a body $(b)$ in the system. 
However, both the interaction  term ${\cal H}_{mn} $ and the total 
solution for the metric tensor $g_{mn}$ in the coordinates $(y^p_a)$ 
appear   to be `parameterized' by the arbitrary functions 
${\cal K}_a, {\cal L}_a, {\cal Q}^\alpha_a$. 
This result  reflects the covariance of the gravitational field equations
as well as the well defined  transformation 
properties of the gauge conditions Eq.(16) used to derive the total solution.
This arbitrariness  suggests that one could choose any form of these functions 
in order to describe the dynamics of the 
extended bodies in the system. However, as we noted earlier, a 
disadvantageous choice of the proper reference frame  (or, equivalently, the 
functions ${\cal K}_a, {\cal L}_a$ and ${\cal Q}^\alpha_a$) might cause an 
unreasonable complication  in the 
future physical interpretations of the results obtained.  

\subsection{Principles of Construction  of the Proper Reference Frame.} 
 
In this Section we will   present a  
way to find the  transformation functions necessary for constructing a 
proper reference frame with well defined properties. 
As one can see from the expressions (12), in the  {\small WFSMA}   
the main contribution to the geometrical properties of the proper reference frame   
in the body's immediate vicinity   comes from its own gravitational field 
$h^{(0)a}_{mn}$. Then, based on the Principle of Equivalence, the   
external gravitational influence should vanish at least to 
first order in the spatial coordinates (\cite{Syn60}, \cite{Man63}).  
The  proper reference frame, constructed this way, should 
resemble the properties of a quasi-inertial (or Lorentzian) reference frame and, 
as such,  will be well suited for discussing the physical experiments. 
Note  that the   tensors $h^{(0)b}_{mn}$ and ${\cal H}_{mn}$ represent 
the real gravitational field which no coordinate transformation 
can eliminate everywhere in the system.
In the case of a massive monopole body, one can eliminate the influence of  
the  external field on the  body's world-line only. However, for an 
arbitrarily shaped extended body, the coupling of the body's intrinsic  multipole
moments to the surrounding gravitational field    changes the physical picture 
significantly. This means that  the definition of the proper reference frame for
the extended body must take  into account this non-linear gravitational coupling. 

In order to suggest the   procedure for the choice of the 
coordinate transformations to the  proper reference frame with well established 
properties, let us 
discuss the general structure of the solution $g^a_{mn}(y^p_a)$ given by 
expression (12). Thus, in the expressions for  
$g^a_{mn}$ one may easily separate the  four physically different terms. 
These terms are:
 
\begin{itemize}

\item[(i).]    The Riemann-flat contribution of the field of 
inertia $\phi^a_{mn} $ given by expression $(16b)$. 
\item[(ii).]   The contribution   of the body's own gravitational field 
$h^{(0)a}_{mn}$. 
 
\item[(iii).]  The term  due to the non-linear interaction of the proper 
gravitational field with an   
 external  field. This contribution is due to 
the  Newtonian potential  and the potential $\Phi_{2}$ in the 
expressions (3). These interaction terms   show  up as the   
coupling  of the body's intrinsic multipole moments with the external 
field.  
 
\item[(iv).] The term  describing the field of  the 
external sources of gravity. This term  comes from the 
transformed solutions $h^{(0)b}_{mn}$
and the interaction term ${\cal H}_{mn}$. 

\end{itemize} 

\noindent The  first contribution depends on the external field   in the 
gravitational domain occupied by  the body $(a)$ and 
appears to be `parametrized'  by the 
transformation  functions Eq.(13). Note that for any choice of these functions, 
the obtained  metric  
$g^a_{mn}$, by the way it was constructed, satisfies  the gravitational field 
equations of the 
specific metric theory of  gravity under study.  
Furthermore, based on the   properties of the proper reference frame   discussed 
above, one may   expect that the functions ${\cal K}_a, {\cal L}_a$ and 
${\cal Q}^\alpha_a$ should form   a background  Riemann-flat inertial space-time 
$\eta_{mn}^a$ in this reference frame  which will  be   tangent 
to the total gravitational field in the vicinity of 
body  $(a)$'s  world-line $\gamma_a$. Moreover, the  
difference of these fields should vanish to  first order with respect to the 
spatial coordinates ({\it i.e.} the `external' dipole moment equals zero
\cite{Tho85}). These conditions, applied to  
moving test particles, are known as Fermi 
conditions (\cite{Mis73}, \cite{Man63}, \cite{Fer22}).  
 We have  extended the applicability of these conditions  
to the case of a system composed of {\small N} arbitrarily shaped 
extended celestial bodies.

In order to obtain the  functions ${\cal K}_a, {\cal L}_a$ and ${\cal Q}^\alpha_a$ 
for the coordinate transformation  Eq.(13) we will 
introduce an iterative procedure which will be based on  a
 multipole power expansion with respect to the unperturbed spherical 
harmonics.  To demonstrate the use of these conditions,  
let us denote $H^a_{mn}(y^p_a)$   
as  the {\it local} gravitational field, {\it i.e.} the field 
which is formed from contributions (ii) and (iii) above. 
The     metric tensor in the {\it local} region 
in this case can be represented by the expression: 
$g^{(loc)}_{mn}(y^p_a)= \eta^{(0)}_{mn}+ H^a_{mn}(y^p_a)$.
 Then the generalized Fermi conditions in the {\it local} 
region of body $(a)$ (or in the immediate vicinity of 
its world-line $\gamma_a$) may be imposed on this {\it local} metric 
tensor   by the following equations:
{} 
$$ \lim_{\gamma\rightarrow \gamma_a} g_{mn}(y^p_a)     =  
\hskip 2mm g^{(loc)}_{mn}(y^p_a) \Big|_{\gamma_a},
 \eqno(18a)$$ 
$$ \lim_{\gamma\rightarrow \gamma_a}\Gamma^{k}_{mn}(y^p_a)    = 
\hskip 2mm \Gamma^{k(loc)}_{mn}(y^p_a) \Big|_{\gamma_a},
   \eqno(18b)$$
\noindent where $\gamma$ is the world-line of the point of interest and 
the quantities $\Gamma^{k(loc)}_{mn}(y^p_a)$  are the  Christoffel 
symbols calculated with respect to the {\it local} gravitational 
field $g^{(loc)}_{mn}(y^p_a)$.
Application of these conditions will  determine the  
 functions ${\cal K}_a, {\cal L}_a, {\cal Q}^\alpha_a$  which are as yet unknown. 
Moreover, this procedure will enable us to derive the second-order 
ordinary differential equations for the functions 
$y^\alpha_{a_0} (y^0_a)$ 
and  ${\cal Q}^\alpha_a (y^0_a, 0)$, or, in other words, to determine the 
equations of the perturbed motion of the center of the 
{\it local} field in the vicinity of body $(a)$.

The relations Eq.(18)  summarize our expectations  
based on the Equivalence Principle 
about the {\it local} gravitational environment of the 
self-gravitating bodies.
By making use of these equations, we will be able  to separate  
the {\it local} gravitational field from the external field   
in the immediate vicinity of the bodies.
However, these conditions  only allow  us to determine the transformation
functions for the free-falling massive monopoles  ({\it i.e.} only up to 
the second order with respect to the spatial coordinates). The transformation 
 functions Eq.(13) in this case will depend only on the unperturbed 
contributions of the external gravitational potentials $U_b$ and $V^\alpha_b$ 
and  their first  derivatives taken on the world-line of   body $(a)$.
The results obtained will not   account for the  contribution of the 
multipolar interaction of the proper gravity with the external field in the 
volume of the extended body. 
This accuracy is sufficient for taking into 
account  the terms describing  the interaction of the   
intrinsic quadrupole moments of the bodies  with the surrounding gravitational 
field, but  some more general  condition, in addition to Eq.(18), must be applied
  in order to account for the higher multipole structure of the bodies.

The conditions  Eq.(18), however,   enable  
one  to obtain the complete solution for the Newtonian function ${\cal K}_a$.
Functions ${\cal L}_a$ and ${\cal Q}^\alpha_a$ may be defined up to the second 
order with respect to the spatial point separation,
 namely  ${\cal L}_a, {\cal Q}^\alpha_a \sim O(|y^\alpha_a|^3) $, so the 
arbitrariness of  higher orders $(k \geq 3)$
in the spatial point separation will remain  in the transformation. 
In order to get the corrections to these functions up to  
$k^{th}$ order $(k \geq 3)$ with respect to the powers of 
the spatial coordinate $y^\lambda_a$, 
one should use conditions which   contain the spatial 
derivatives of the metric tensor to order $(k-1)$. 
The mathematical methods of  modern theoretical physics  
generally consider  {\it local} geometrical quantities only and  
involve second order differential equations.
These equations alone  may not be very helpful  for 
constructing the remaining terms in functions ${\cal L}_a, {\cal Q}^\alpha_a$ 
up to the order $k \geq 2$. 
However, following Synge \cite{Syn60}, one may   apply additional geometrical 
constructions,  
such as properties of  the Riemann  tensor and  the Fermi-Walker 
transport law (\cite{Ni78}, \cite{Man63}-\cite{Li79b}). 
Another possibility is to   postulate   the existence of 
so called   `external multipole moments'  (\cite{Bru88a}, \cite{Dam91}, 
\cite{Tho80},  \cite{Bla86}). However, those 
moments are defined through 
vacuum solutions of the Hilbert-Einstein field equations
of general  relativity in an inertial reference frame, while
 the influence of  external sources of gravity 
are ignored. The fact of defining the moments in this way is essentially 
equivalent to defining the structure of the proper reference frame
 for the body under question. 

The most natural approach to define the desirable 
  properties of the proper quasi-inertial reference frames for the 
system of extended and deformable bodies  is to   
study the  motion of this system in an 
arbitrary ${\cal KLQ}$-parametrized frame. There exist two 
different ways to do this, namely: (i) to study the infinitesimal 
motion  of each element of the body, or (ii) to study the motion of a 
whole body with respect to an accelerated frame  attached, say, to the center of 
inertia of the {\it local} fields of matter, inertia, and   gravity. In our
 method we will use the second way and  will study the dynamics of the 
body in its own reference frame. Our analysis will be directed toward finding   
the functions ${\cal K}_a, {\cal L}_a$ and ${\cal Q}^\alpha_a$ with the condition 
 that the Riemann-flat inertial space-time $\eta^a_{mn}(y^p_a)$ corresponding to 
these functions will be tangent to the total Riemann metric $g_{mn}(y^p_a)$ of 
the entire system in the body's vicinity.
Physically, one expects that this inertial space-time will produce 
a `fictitious' (or inertial)   force  with the density ${\vec f}_{\cal KLQ}$
acting on the body in its   proper reference frame. At the same 
time, the body  is under influence of the  
overall   real force due to the {\it local} fields of   matter and 
gravity with the density ${\vec f}_0$. Thus, the condition for finding the 
transformation functions ${\cal K}_a, {\cal L}_a$ and ${\cal Q}^\alpha_a$  is 
conceptually simple: the difference between these two densities 
${\vec {\cal F}}={\vec f}_0-{\vec f}_{\cal KLQ}$ should vanish 
after integration (or averaging) over the body's compact   volume:  
{}
$$ \delta {\vec  F} = \int_a d^3y'_a{\vec {\cal F}}=
\int_a d^3y'_a\Big( {\vec f}_0-{\vec f}_{\cal KLQ}\Big)=0. \eqno(19)$$
 
\noindent Note that the notion of `the center of mass' in this case  
loses its practical value, and one  should substitute instead
`the {\it local}  center of inertia'.  Thus, the  force ${\vec f}_{\cal KLQ}$ 
should provide the 
overall static equilibrium for the body under consideration in the
{\it local}  center of inertia, which is defined for all three  fields present
in the immediate vicinity of the body, namely:  matter, inertia and gravity.     
Let us mention here that in practice it is not possible to separate these 
two forces ${\vec f}_0$ and ${\vec f}_{\cal KLQ}$ from each other. Fortunately, 
it is possible to obtain  at once the difference between them ${\vec {\cal F}}$. 
This will considerably simplify  the further analysis \cite{Tur96}.

In order to construct the necessary solution 
for the functions  ${\cal K}_a, {\cal L}_a$ and ${\cal Q}^\alpha_a$ 
in a way  that will be valid for a wide class of    
metric theories of gravity, one must first  analyze 
 the conservation laws in an arbitrary 
${\cal KLQ}$-parameterized reference frame. This could be 
done based on the conservation law 
for the density of the total energy-momentum
tensor $\hat{T}^{mn}$ of the whole isolated  {\small N}-body system:
{}
$$\nabla^a_n\widehat{T}^{mn} (y^p_a) = 0, \eqno(20)$$
 \noindent where $\nabla^a_n$ is the covariant derivative with
 respect to total Riemann  metric $g^a_{mn} (y^p_a)$ in these coordinates.
Then, by using a standard technique of integration with Killing vectors,
 one will have to integrate this equation 
over the  compact volume of the body $(a)$, and 
 one can  obtain  the equations of motion of 
the extended body (\cite{Wil93}, \cite{Foc57}, \cite{Chandra65}). 
Then the  necessary conditions, equivalent to those of Eq.(18), 
 may be formulated as the   requirement that the 
translational motion of the extended bodies vanish in their own   reference 
frames.  This  corresponds  to the following conditions    applied to 
the  dipole mass moment  ${\vec m}_a\equiv I^{\{1\}}_a$:
{}
$${d^2{\vec m}_a\over d{y^0_a}^2}={d{\vec m}_a\over dy^0_a}
 ={\vec m}_a= 0,\eqno(21a)$$
\noindent where the quantity  ${\vec m}_a$ is calculated based on the total
energy-momentum tensor matter, inertia and gravitational 
field taken jointly (similar to condition of Eq.(4)).  
These conditions may also be presented in a different form. 
Indeed, if we  require that the 
total momentum ${\vec P}_a$ of the {\it local} fields of matter, inertia 
and gravity in the vicinity of the 
extended body vanish, we will have the following   physically equivalent 
condition:
{}
$$ {d{\vec P}_a\over dy^0_a}= {\vec P}_a=0. \eqno(21b)$$
\noindent These conditions  finalize the  formulation of the basic principles of 
construction of the relativistic theory of   celestial   reference frames
in the {\small WFSMA}. 
In the next Section we will present the results  obtained when 
these principles are applied.

\section{Properties of the  Proper Reference Frame.}
In this Section  we will present the basic results obtained for
the relativistic coordinate transformations in the  {\small WFSMA}.
We will present the results for the transformation functions Eq.(13) as well as 
the solution  for the metric tensor in the coordinates of the proper reference 
frame.

\subsection{Coordinate Transformations and Metric Tensor.}

By taking into account all the conditions presented  in the previous Section, 
  one can   obtain  the set of differential equations on the transformation  
functions ${\cal K}_a, {\cal Q}^\alpha_a$ 
and ${\cal L}_a$. The solutions to these equations  can be given    as   follows:
{}
$${\cal K}_a(y^0_a,y^\nu_a)  =\int^{y^0_a} \hskip -10pt  dt'  
\Big( \sum_{b\not=a} \Big< U_b\Big>_a- 
{1\over2}{v_{a_0}}_\nu v_{a_0}^\nu\Big)  
-{v_{a_0}}_\nu\cdot y^\nu_a +  {\cal O}(c^{-4})y^0_a, \eqno(22a)$$
{}
$${\cal Q}^\alpha_a(y^0_a,y^\nu_a) = - \gamma \sum_{b\not=a} \Big( y^\alpha_a 
 y^\beta_a \cdot \Big<\partial_\beta U_b \Big>_a-
{1\over2} {y_a}_\beta  y^\beta_a  
\Big<\partial^\alpha U_b\Big>_a  
 + y^\alpha_a  \Big<U_b\Big>_a \Big)+$$
{}
$$+{y_a}_\beta  \int^{y^0_a} \hskip -10pt dt' 
\Big( {1\over2}a^{[\alpha}_{a_0}  v^{\beta]}_{a_0}  + 
(\gamma+1) \sum_{b\not=a}\Big[
\Big<\partial^{[\alpha} V^{\beta]}_b\Big>_a   
+\Big<\partial^{[\alpha} U_b v^{\beta]}\Big>_a \Big]\Big)- $$
{}
$$- {1\over2} v^\alpha_{a_0} v^\beta_{a_0}{y_a}_\beta  
 + w^\alpha_a(y^0_a) +\sum^{k}_{l \ge 3} {Q^\alpha_a}_{\{L\}} (y^0_A) \cdot 
y^{\{L\}}_a+{\cal O}(|y^\nu_a|^{k+1})+ 
{\cal O} (c^{-4}) y^\alpha_a,   \eqno(22b)$$ 
{}
$${\cal L}_a(y^0_a,y^\nu_a) =  \sum_{b\not=a} \Big( {1\over2}\gamma\hskip 2pt
y_{a\beta} y^\beta_a \cdot {\partial  \over\partial y^0_a}\Big<U_b\Big>_a  - 
(\gamma+1)\hskip 2pt y^\lambda_a  y^\beta_a \cdot 
\Big[\Big< \partial_\lambda {V_b}_\beta \Big>_a   
+\Big<v_\beta \partial_\lambda U_b \Big>_a \Big] +$$
{}
$$+ \gamma \hskip 2pt{v_{a_0}}_\beta \Big[y^\beta_a  y^\lambda_a \cdot 
\Big< \partial_\lambda  U_b \Big>_a - 
{1\over2} y_{a_\lambda} y^\lambda_a \cdot 
\Big< \partial^\beta  U_b \Big>_a \Big] \Big) +$$
{}
$$+ y_{a_\lambda}{v_{a_0}}_\beta \int^{y^0_a} \hskip -10pt dt'
 \Big( {1\over2} a^{[\lambda}_{a_0} v_{a_0}^{\beta]}+ 
(\gamma+1) \hskip 2pt\sum_{b\not=a}
\Big[\Big<\partial^{[\lambda} V^{\beta]}_b  \Big>_a   
+\Big<v^{[\beta}\partial^{\lambda]} U_b \Big>_a\Big]\Big)+$$
{}
$$ + y_{a_\beta} \Big[ (\gamma+1)\hskip 2pt v^\beta_{a_0}\sum_{b\not=a} 
\Big<U_b\Big>_a  
 -  2(\gamma+1)\hskip 2pt\sum_{b\not=a} \Big<V^\beta_b\Big>_a  
 - \dot {w}^\beta_{a_0}(y^0_a)  \Big] - $$
{}
$$ - \int^{y^0_a} \hskip -10pt dt' \Big[ \sum_{b\not=a}
\Big<W_b\Big>_a  + {1\over2}\Big(\sum_{b\not=a} \Big< U_b\Big>_a  - 
 {1\over2}{v_{a_0}}_\beta v^\beta_{a_0}\Big)^2 
  + {v_{a_0}}_\mu\dot{w}^\mu_{a_0}(t')\Big]+$$
{}
$$+ \sum^{k}_{l\ge3}{L_a} _{\{L\}}(y^0_a) \cdot y^{\{L\}}_a+
  {\cal O}(|y^\nu_a|^{k+1}) +{\cal O}(c^{-6}),\eqno(22c)$$ 
\noindent  where the dot over the function $w^\mu_{a_0}$ denotes the regular 
time derivative.
Note that in the case of the free-falling  structureless  test
particle with conserved mass density given by
${\hat \rho}_a(y^0_a,y^\nu_a)=m_a\delta(y^\nu_a)$, the  
functions Eq.(22)  correspond  to the coordinate transformations 
to   the proper reference frame  defined on the 
geodesic world line of this particle.  
The time-dependent functions ${Q^\alpha_a} _{\{L\}}$
and ${L_a} _{\{L\}}$ in the expressions Eq.(22) are the 
contributions coming from  the higher multipoles $(l\ge3)$
(both mass and current induced  ones) 
of the external gravitational field generated by the 
 bodies $(b\not=a)$  in the system.
These functions  enable one to take into account the geometrical features of 
the proper reference frame  with respect to three-dimensional  
spatial rotation. The form of these functions may be chosen  arbitrarily. This 
freedom enables one to choose any coordinate dependence for the terms with 
$l\ge3$ in order to  describe  the motion of the highest monopoles. Moreover, 
one may show that, even though the total solution to the metric tensor 
$g_{mn}(x^p)$ in the  barycentric inertial  reference frame  
 resembles the form of the one-body solution Eq.(3), if one will expresses  
this solution   through the proper multipole moments of the bodies, it will 
contain the contributions from the functions ${Q^\alpha_a} _{\{L\}}(y^0_a)$ 
and ${L_a} _{\{L\}}(y^0_a)$.

Within the accuracy  necessary for future analysis, 
we present the equations for both   time-dependent functions 
$y_{a_0}^\alpha$ and $w_{a_0}^\alpha$ written with
respect to  time $y^0_a$ of the proper reference frame.
Thus the  Newtonian acceleration of body $(a)$ with respect to 
the barycentric reference frame  may be  described as follows
{}
$$a^\alpha_{a_0}(y^0_a)=-\eta^{\alpha\mu}\sum_{b\not=a}
\Big<{\partial  U_b\over \partial y^\mu_a}\Big>_a+{\cal O}(c^{-4})=
-\eta^{\alpha\mu}\sum_{b\not=a}{1\over m_a}\int_a d^3y'^\nu_a
\hskip 2pt{\hat\rho}_a(y'^p_a)
{\partial  U_b(y'^p_a)\over \partial y'^\mu_a}+{\cal O}(c^{-4}).\eqno(23a)$$
\noindent Whereas the  post-Newtonian part of the acceleration may be 
represented by the expression 
{}
$$\ddot{w}^\alpha_{a_0}(y^0_a)=
\sum_{b\not=a} \Big(\eta^{\alpha\mu}\Big<{\partial W_b\over \partial 
y^\mu_a}\Big>_a+ v^\alpha_{a_0}{\partial \over \partial y^0_a} \Big< U_b\Big>_a  -
  2(\gamma+1)\hskip 2pt{\partial \over \partial y^0_a}
\Big< V^\alpha_b\Big>_a\Big)-$$
{}
$$- {1\over2}v^\alpha_{a_0}{v_{a_0}}_\beta a^\beta_{a_0} + 
\gamma \hskip 2pt a^\alpha_{a_0}\sum_{b\not=a} \Big<U_b\Big>_a  
-\Big<\partial^\alpha {\overline U} \Big(\Pi-{1\over 2}(1+2\gamma) v_\mu v^\mu+
{3p\over {\overline\rho}}+(2\beta+2\gamma-1){\overline U}\Big)\Big>_a+ $$
{}
$$+{a_{a_0}}_\beta \int^{y^0_a} \hskip -10pt dt' 
\Big( {1\over2}a^{[\alpha}_{a_0}  v^{\beta]}_{a_0}  + 
(\gamma+1) \sum_{b\not=a}\Big[
\Big<\partial^{[\alpha} V^{\beta]}_b\Big>_a   
+\Big<v^{[\beta}\partial^{\alpha]} U_b \Big>_a \Big]\Big) -$$
{}
$$-{1\over 5}{\ddot{a}_{a_0}}{}_\mu \int_a d^3 y'^\nu_a{\hat \rho}_a
\Big(y'^\alpha_a y'^\mu_a+{1\over2}\gamma^{\alpha\mu}y'_{a_\lambda} y_a'^\lambda 
 \Big)   -{1\over \pi m_A} \oint_A d S^\beta_A 
\Bigg(\partial ^\alpha  {\overline U}{\partial  {\overline V}_\beta \over 
\partial y^0_A}+\partial _\beta    {\overline U}{\partial   
{\overline V}^\alpha\over \partial y^0_A} -$$
{}
$$-\delta^\alpha_\beta \partial_\lambda{\overline U}
 {\partial {\overline V}^\alpha\over \partial y^0_A}+ 
\partial^{[\nu}{\overline V}^{\alpha]}\partial_{[\nu}{\overline V}_{\beta]}-
{1\over2}\delta^\alpha_\beta \partial^{[\mu}
{\overline V}^{\nu]}\partial_{[\mu}{\overline V}_{\nu]} 
+ {1\over 2} \delta^\alpha_\beta\Big({\partial 
{\overline U}\over \partial y^0_A}\Big)^2\Bigg)$$ 
{}
$$-\int_a d^3 y'^\nu_a \Bigg({\hat \rho}_a  
\sum^k_{l\ge3}\Big[\partial^2_{00}{Q^\alpha_a}_{\{L\}}(y^0_a)+
{a_{a_0}}_\lambda {Q^\lambda_a}_{\{L\}}(y^0_a)\partial^\alpha\Big] y^{\{L\}}_a 
 +2 {\hat \rho}_a  v^\mu\sum^k_{l\ge3}\partial_0
{Q^\alpha_a}_{\{L\}}(y^0_a)\partial_\mu y^{\{L\}}_a+$$
{}
$$+( {\hat \rho}_a  v^\mu v^\lambda-\gamma^{\mu\lambda}p)
\sum^k_{l\ge3}{Q^\alpha_a}_{\{L\}}(y^0_a)\partial^2_{\mu\lambda} y^{\{L\}}_a- 
{\hat \rho}_a  {\partial\overline{U}(y^p_a)\over \partial y^\mu_a}
\sum^k_{l\ge3}{Q^\alpha_a}{}_{\{L\}}(y^0_a) \partial^\mu y^{\{L\}}_a
\Bigg)
+ {\cal O}(c^{-6}). \eqno(23b)$$
\noindent 
For practical purposes one may find the value of the surface integrals 
in the expression Eq.$(23b)$ by  performing  an iteration procedure.    
It can be  shown that the lowest multipole moments of the bodies 
will not contribute to  this surface integration. However, the 
general results  will fully depend   on the non-linear 
interaction of the intrinsic multipole moments with the external 
gravity in the {\it local} region at the vicinity of the body under 
consideration. This additional iterative option   will make  all 
the  results obtained with the 
proposed formalism  easy to use in  practical applications.   

The  expressions Eq.(23) are the two parts of the 
force  necessary   to keep the body $(a)$ in it's orbit (world tube)
in the  {\small N}-body system. 
These expressions are written in terms of proper time, and  if one performs 
the coordinate transformation from the coordinates $(y^p_a)$   
to those   of   $(x^p)$ for
 all the functions and potentials  entering both equations  (23),
and takes into account the lowest intrinsic multipole moments of the bodies 
only, one  obtains the simplified equations of motion for the extended  bodies
Eqs.$(B1$-$B7)$ written in the coordinates $(x^p)$ of the barycentric inertial 
reference frame. 

The transformations Eqs.(13), (22)-(23) are the generalization of the Poincare' 
group for the case of motion of  extended  self-gravitating bodies. Note,  that  
Chandrasekhar and Contopulos  \cite{Chandra67} obtained the coordinate 
transformations with the limitation of not 
violating the form-invariancy of the metric tensor. This ensured   that the 
equations of motion would  preserve their form. However, in 
our case, the transformation  of coordinates to the proper reference frame  
connected with body $(a)$,  were obtained with the generalized  
Fermi conditions Eq.(18) and Eq.(21), and hence they changed the metric tensor 
considerably. The most notable  contribution to the 
equations of motion of the test particle  orbiting this body comes from
the gravitational field of  body $(a)$ itself.  The influence of the 
external sources of gravity presents itself in the form of the tidal terms only. 
This can be seen from the metric tensor $g^a_{mn}$ in the coordinates $(y^p_a)$,  
which may be obtained  in the following form:
{}
$$g^a_{00}(y^p_a) =1-2{\overline U}+ 2{\overline W} +  
y^\mu_a y^\beta_a \cdot \Big[ \gamma \hskip 2pt\eta_{\mu\beta} 
\hskip 1mm {a_{a_0}}_\lambda a^\lambda_{a_0} -  
(2\gamma-1)\hskip 2pt{a_{a_0}}_\mu {a_{a_0}}_\beta + $$
{}
$$+ \sum_{b\not=a}
{\partial\over\partial y^0_a}\Big(\gamma \hskip 2pt\eta_{\mu\beta} 
{\partial\over\partial y^0_a}\Big< U_b\Big>_a-
(\gamma+1)\Big[\Big<\partial_{(\mu} {V_b}_{\beta)}\Big>_a   
+\Big<v_{(\beta}\partial_{\mu)} U_b\Big>_a \Big]\Big) +$$
{}
$$ + 2  \sum^{k}_{l\ge3} 
 \Big[ \partial_0 {L_a} _{\{L\}}(y^0_a)
 + {v_{a_0}}_\beta \partial_0 Q^\beta_{a\{L\}} (y^0_a)\Big]\cdot
 y^{\{L\}}_a 
  +  
{\cal O}(|y^\nu_a|^{k+1}) +{\cal O}(c^{-6}), \eqno(24a) $$ 
{}
$$g^{a}_{0\alpha}(y^p_a) = 2(\gamma+1)\hskip2pt \eta_{\alpha\epsilon}
{\overline V^\epsilon}+{1\over2} \gamma\hskip2pt\Big( 
\eta_{\alpha(\epsilon}\delta^\beta_{\lambda)} -
\eta_{\epsilon\lambda} \delta^\beta_\alpha\Big) \hskip 1mm 
y^\epsilon_a y^\lambda_a \cdot 
{\dot a}_{a_0}{}_\beta  + $$
{}
$$+\sum^{k}_{l \ge 3} \Big[\eta_{\alpha \lambda}
 \partial_0 {Q^\lambda_a} _{\{L\}} (y^0_a) + 
\Big({L_a} _{\{L\}}(y^0_a)  + {v_{a_0}}_\beta \cdot 
{Q^\beta_a} _{\{L\}} (y^0_a)\Big) 
{\partial \over \partial y^\alpha_a}\Big]
\cdot  y^{\{L\}}_a  + $$
{}
$$+{\cal O}(|y^\nu_A|^{k+1})+  {\cal O}(c^{-5}), \eqno(24b)$$
{}
$$g^{a}_{\alpha\beta}(y^p_a) = 
\eta_{\alpha\beta}\Big(1 + 2\gamma{\overline U}\Big) +$$
{}
$$+\sum^{k}_{l \ge 3} \Big[ \eta_{\alpha\lambda} {Q^\lambda_a}_
{\{L\}} (y^0_a)  
{\partial \over \partial y^{\beta}_a}  +
\eta_{\beta\lambda} {Q^\lambda_a}_
{\{L\}} (y^0_a)  {\partial \over \partial y^\alpha_a}
\Big]\cdot  y^{\{L\}}_a + {\cal O}(|y^\nu_a|^{k+1})+ 
{\cal O}(c^{-4}), \eqno(24c)$$

\noindent where  the total gravitational  potential ${\overline U}$  
at the vicinity of body $(a)$ is composed of the local Newtonian 
potential generated by the body $(a)$ itself and the tidal gravitational 
potential produced by  external   sources of gravity:
{} 
$$\overline{U}(y^p_a)=U_a(y^p_a)+
\sum_{b\not=a} U_b(y^q_b(y^p_a))-{\partial {\cal K}_a(y^p_a)\over 
\partial y^0_a} - {1\over2}v^\mu_{a_0}{v_{a_0}}_\mu=$$
$$=\sum_b U_b\big(y^q_b(y^p_a)\big)-\sum_{b\not=a}
\Big(y^\beta_a\Big<{\partial U_b\over \partial y^\beta_a}\Big>_a +
\Big<U_b\Big>_a \Big)+ {\cal O}(c^{-4}), \eqno(25) $$
\noindent with the  Newtonian acceleration of  body $(a)$ 
given by Eq.$(23a)$. This potential is the solution of the 
Poisson equation in coordinates $(y^p_a)$, namely:
{}
$$  \eta^{\mu\lambda}{\partial^2\over \partial y^\mu_a \partial y^\lambda_a} 
{\overline U}(y^p_a)= 4\pi\sum_b{\hat \rho}_b(y^p_a)
\big(1+{\cal O}(c^{-2})\big). \eqno(26a)$$
\noindent By satisfying the   requirement  that the body   is 
 at rest in its proper reference frame,  we derive  
 the following results:
{}
$$\Big<{\overline U}\Big>_a =
\int_a d^3y'^\nu_a\hskip 2pt{\hat\rho}_a(y'^p_a){\overline U}(y'^p_a)=
\int_a d^3y'^\nu_a\hskip 2pt{\hat\rho}_a(y'^p_a)U_a(y'^p_a)=2E_a, \eqno(26b)$$ 
{}
$$\Big<{\partial  {\overline U}\over \partial y^\mu_a}\Big>_a 
=\int_a d^3y'^\nu_a\hskip 2pt{\hat\rho}_a(y'^p_a)
{\partial  {\overline U}(y'^p_a)\over \partial y'^\mu_a}=0.$$ 
 
The quantity  ${\overline V}^\alpha(y^p_a)$ in the expressions Eq.(24) 
is the total vector-potential
produced by all the bodies in the system represented in the 
coordinates $(y^p_a)$ of the reference frame:
{}
$${\overline V}^\alpha(y^p_a)=\sum_b V^\alpha_b(y^q_b(y^p_a))-\sum_{b\not=a}
\Big(y^\mu_a \Big[\Big<\partial_\mu V^{\alpha}_b \Big>_a+ 
\Big<{\underline v}^\alpha\partial_\mu  U_b \Big>_a\Big]+\Big<
V^\alpha_b\Big>_a\Big)+ $$
{}
$$+{1\over10}(3y^\alpha_a y^\lambda_a-\eta^{\alpha\lambda}{y_a}_\mu y^\mu_a\Big)
\dot{a}_{a_0}{}_\lambda+{\cal O}(c^{-4}). \eqno(27) $$
\noindent  This potential  satisfies the   equation
{}
$$ \eta^{\mu\lambda} {\partial^2\over \partial y^\mu_a\partial y^\lambda_a} 
{\overline V}^\alpha(y^p_a)= -4\pi\sum_b{\hat \rho}_b(y^p_a)
v^\alpha(y^p_a)\big(1+{\cal O}(c^{-2})\big). \eqno(28)$$
\noindent The form  of the solution for ${\overline V}^\alpha$ Eq.(27) is  
chosen in such a way that it  satisfies the Newtonian equation of continuity  
written in the proper reference frame: $ {\partial_0}{\overline\rho} (y^p_a)+
{\partial_\mu} \big[{\overline\rho} (y^p_a)v^\mu(y^p_a) \big] 
={\cal O}(c^{-3})$, 
which provides the following Newtonian-like relation between 
these quantities:
{}
$$ {\partial  {\overline U} \over \partial y^0_a}=
{\partial  {\overline V}^\mu \over \partial y^\mu_a}.$$

Another quantity  we have introduced in the formulae Eq.(24)  is 
${\overline W(y^p_a)}$. This is the post-Newtonian 
contribution to the component $g_{00}$ of the effective metric tensor in the 
coordinates $(y^p_a)$ of the proper reference frame  and it is given as:
{}
$${\overline W(y^p_a)}=\sum_b W_b(y^q_b(y^p_a))-\sum_{b\not=a}
\Big(y^\mu_a \Big<{\partial W_b\over \partial y^\mu_a}\Big>_a +
\Big<W_b\Big>_a\Big) + {\cal O}(c^{-6}), \eqno(29) $$
\noindent  with the expressions for 
the functions $W_a$ and $W_b$   given as follows:
{}
$$W_a(y^p_a) = \beta \hskip 2pt U^2_a(y^p_a) +\Psi_a(y^p_a) +
2 a^\lambda_{a_0}\cdot{\partial \over 
\partial y^\lambda_a}\chi_a (y^p_a)+{1\over2}{\partial^2\over\partial 
{y^0_a}^2} \chi_a(y^p_a)+$$
{} 
$$+\sum_{b\not=a}\Bigg(2\beta \hskip 2pt U_{a}(y^p_a) U_b(y^p_a)  
-(3\gamma+1-2\beta)\int_a{d^3y'^\nu_a \over{|y^\nu_a - y'^\nu_a|}} 
 \rho_a(y^0_a,y'^\nu_a) U_b(y^0_a,y'^\nu_a)\Bigg)+$$
{}
 $$+ \sum^{k}_{l\ge3}{Q^\lambda_a} _{\{L\}} (y^0_a) 
 \int_a d^3y_a'^\nu \rho_a (y_a^0,y_a'^\nu)
  {\partial \over\partial y'^\lambda_a} 
\Big[{ y^{\{L\}}_a-y'^{\{L\}}_a \over{|y_a^\nu - y_a'^\nu|}}\Big]+ 
O(|y^\nu_a|^{k+1})+ {\cal O}(c^{-6}). \eqno(30a)$$ 
{} 
$$W_b(y^p_a) = \beta U_b(y^p_a)\sum_{c\not=a} U_c(y^p_a) +\Psi_b(y^p_a)  +
2 a^\lambda_{a_0}\cdot 
{\partial \over \partial y^\lambda_a} \chi_b (y^p_a)+{1\over2}
{\partial^2\over\partial {y^0_a}^2} \chi_b(y^p_a) - $$
{}
$$-(3\gamma+1-2\beta)\int_b{d^3y'^\nu_a \over{|y^\nu_a-y'^\nu_a|}} 
\rho_b(y^0_a, y'^\nu_a)   \sum_{b'}U_{b'}(y^0_a,y'^\nu_a)+  $$
{}
$$+ \sum^k_{l\ge3}{Q^\lambda_a} _{\{L\}} (y^0_a) \cdot 
\int_bd^3y_a'^\nu \rho_b\Big(y_a^0, y_a'^\nu+y_{ba_0}^\nu(y_a^0)\Big) 
{\partial \over\partial y'^\lambda_a} 
 \Big[{ y^{\{L\}}_a-y'^{\{L\}}_a \over{|y_a^\nu - y_a'^\nu|}}\Big] +$$
{}
$$+ O(|y^\nu_a|^{k+1})+ {\cal O}(c^{-6}). \eqno(30b)$$
\noindent  The  functions $W_a$ and $W_b$ fully represent the non-linearity of 
the total post-Newtonian gravitational  field in the proper reference frame. 
These functions contain the contributions of two sorts: (i) the gravitational 
field produced by the external $(b\not=a)$ bodies in the system, and (ii) the 
field of inertia caused by the accelerated (the terms with $a^\mu_{a_0}$) and 
non-geodesic motion (due to the coupling of the proper multipole moments of the 
body $(a)$  with  the external gravitational field and the self-action 
contributions both  given by the terms with ${Q^\lambda_a} _{\{L\}}$ in the  
expressions Eq.(30)) of the proper reference frame.

\subsection{The Fermi Normal Coordinates and the Equations of the Spacecraft 
Motion.}

The expressions  Eq.(24) are the general two-parametric 
solution for the field equations of general relativity and scalar-tensor 
theories of gravity in  {\small WFSMA}. 
These expressions  satisfy  the generalized Fermi conditions 
Eq.(18) at the immediate vicinity of  body $(a)$.
This solution reflects the geometrical features of 
the proper reference frame  with respect to the special properties 
of the motion of the $k$-th multipoles of the extended bodies.
In order to find the  unknown  functions 
${Q^\alpha_a} _{\{L\}}(y^0_a)$ and ${L_a} _{\{L\}}(y^0_a)$ up to the 
$k$-th $\hskip 1mm (k \geq 3)$ order,
one should use the conditions which will contain the spatial 
derivatives from the metric tensor to the $(k-1)$ order. 
 Thus, following Synge \cite{Syn60} in addition to the Fermi conditions 
\cite{Man63}, one may  apply the   Fermi-Walker transport law 
\cite{Mis73}. 
Another possible method is to use the `external' multipole moments 
as defined for the gravitational wave theory (\cite{Tho80}-\cite{Bla89}). 
However, one can show that in
 {\small WFSMA}  the functions 
${Q^\alpha_a} _{\{L\}}(y^0_a)$ and ${L_a} _{\{L\}}(y^0_a)$   
 may be chosen in such a way 
that the metric tensor Eq.(24) in a proper reference frame 
will take  the  form corresponding to any of these multipolar
expansions. One can show that  this happens  because  of the 
`external' multipole moments in the transformation functions, 
and that this  simply corresponds to the 
choice of coordinates of the reference frame  with  specific dynamical 
properties. Thus, for example, one can construct the proper reference frame for
the Fermi normal coordinates \cite{Man63}.
Note that  these conditions require   corrections up to 
the third order with respect to the spatial coordinates  in
 the transformation functions 
${\cal Q}^\alpha_a$ and ${\cal L}_a$ (\cite{Ash86}, \cite{Ash90}, 
\cite{Li79a}-\cite{Mar94}). The necessary   corrections to the functions Eq.(22)
have the following 
form\footnote{ With $\delta^\alpha_\beta$ is being the Kronekker symbol 
and $\eta^{\lambda\mu}\eta_{\lambda\nu}=\delta^\mu_\nu.$}: 
{}
$$ {Q^\alpha_a}_{\nu_3}(y^p_a) = {1\over6}\gamma 
\sum_{b \not=a}y^\mu_a y^\nu_a y^\beta_a\cdot\Big(\eta^{\alpha\lambda} 
\eta_{\mu\nu}\Big<\partial^2_{\beta\lambda} U_b\Big>_a   - 
2 \delta^\alpha_\beta\Big<\partial^2_{\mu\nu} U_b\Big>_a \Big)
 +{\cal O}(|y^\nu_a|^4), \eqno(31a)$$ 
{}
$$ {L_a}_{\nu_3}(y^p_a) = {1\over 6}\sum_{b\not=a}
y^\mu_a y^\nu_a y^\beta_a \cdot\Big( \gamma \hskip 2pt\eta_{\mu\nu} 
{\partial\over\partial y^0_a}\Big<\partial_{\beta} U_b \Big>_a  
 -2(\gamma+1)\hskip 2pt\Big<\partial^2_{\mu\nu} {V_b}_\beta\Big>_a  -$$ 
{}
$$ -\gamma \hskip 2pt{v_{a_0}}_\sigma\Big[\eta^{\sigma\lambda} 
\eta_{\mu\nu}\Big<\partial^2_{\beta\lambda} U_b\Big>_a    - 
2 \delta^\sigma_\beta\Big<\partial^2_{\mu\nu} U_b\Big>_a\Big]  
\Big)+{\cal O}(|y^\nu_a|^4). \eqno(31b)$$ 

\noindent Given this form for the corrections, the   metric tensor Eq.(24)  
 will take the  form corresponding to generalized Fermi normal 
coordinates  chosen in the proper reference frame: 
{}
$$g^{\cal F}_{00}(y^p_a)  = 1-2{\overline U}(y^p_a)+2W_a(y^p_a)+$$
{}
$$+\Bigg( \sum_{b\not=a}\Big[\Big<\partial^2_{\mu\nu}  W_b\Big>_a+
{\partial \over \partial y^0_a}\Big[\gamma \hskip 2pt \eta_{\mu\nu} 
{\partial \over \partial y^0_a}\Big<  U_b\Big>_a-
(\gamma+1)\hskip 2pt \Big<\partial_{(\nu} V_{b\mu)}\Big>_a\Big] \Big] +$$
{}
$$ + \gamma \hskip 2pt \eta_{\mu\nu} \hskip 1mm {a_{a_0}}_\lambda 
a^\lambda_{a_0} -  (2\gamma-1) \hskip 2pt {a_{a_0}}_\mu 
{a_{a_0}}_\nu\Bigg) \cdot y^\mu_a y^\nu_a  +  {\cal O}(c^{-6}) + 
 {\cal O}(|y^\nu_a|^{3}), \eqno(32a) $$
{}
$$g^{\cal F}_{0\alpha}(y^p_a)  = 2(\gamma+1)\hskip 2pt\eta_{\alpha\epsilon} 
V^\epsilon_a(y^p_a)+
{2\over3}\Bigg(\gamma\hskip 2pt\Big(\eta_{\alpha\mu} \dot{a}_{a_0}{}_\nu-  
\eta_{\mu\nu}\dot{a}_{a_0}{}_\alpha\Big) + $$
{}
$$+  (\gamma+1) \sum_{b\not=a} \Big[ \eta_{\alpha\lambda} 
\Big<\partial^2_{\mu\nu}  V^{\lambda}_b\Big>_a  -
\eta_{\nu\lambda}\Big<\partial^2_{\mu\alpha}V^{\lambda}_b\Big>_a
\Big]\Bigg) \cdot y^\mu_a y^\nu_a  
 +  {\cal O}(|y^\nu_a|^{3}) +   {\cal O}(c^{-5}), \eqno(32b)$$ 
{}
$$ g^{\cal F}_{\alpha\beta}(y^p_a)  =   
\eta_{\alpha\beta} \Big(1 + 2 \gamma  U_a(y^p_a)\Big)+
{1\over3} \gamma \sum_{b\not=a}\Big[  \eta_{\alpha\beta} 
\Big<\partial^2_{\mu\nu}U_b\Big>_a    + 
\eta_{\mu\nu} \Big<\partial^2_{\alpha\beta}U_b\Big>_a - $$
{}
$$-\eta_{\beta\mu}\Big<\partial^2_{\alpha\nu} U_b\Big>_a - 
\eta_{\alpha\nu} \Big<\partial^2_{\beta\mu} U_b\Big>_a \Big] 
\cdot y^\mu_a y^\nu_a + {\cal O}(|y^\nu_a|^{3}) +
{\cal O}(c^{-4}), \eqno(32c)$$ 
\noindent with the corresponding equation for $a^\alpha_{a_0}$
given by Eq.$(23a)$.
Thus we have obtained the form of the metric tensor in   Fermi 
normal coordinates and the coordinate transformations 
leading to this form. These transformations are defined up 
to the third order with respect to the spatial coordinates.
This accuracy is sufficient to analyze the motion of a spacecraft 
in orbit around Mercury consistent with  parameterized 
relativistic gravity.
 
We now    obtain the equations of the spacecraft motion in a Hermean-centric 
reference frame. To do  this, we consider a Riemann  space-time whose metric 
coincides with the metric of  {\small N}  moving extended bodies.  We  study 
the motion of a point body in the  neighborhood of body $(a)$. The 
expression for the acceleration of the point body $a^\alpha _{(0)}$ can be 
obtained in two ways: either by using the equations of geodesics of Riemann  
space-time $du^n/ds+\Gamma^n_{mk}u^mu^k=0$ or by computing the acceleration of 
the center of mass of the extended body and then letting all quantities 
characterizing its internal structure and proper gravitational field tend to zero. 
In either case  one obtains the same result.
While the full derivation of the equations of motion is given in the Appendix C, 
we present here the restricted version of the equations $(C4)$ which is 
consistent with the  expected accuracy for ESA's {\it Mercury Orbiter} mission. 
This limited accuracy permits us to completely neglect 
contributions proportional to the spatial coordinates $y^\mu_a$.  
The planeto-centric equations of satellite motion around Mercury can be 
represented  by a series in $1/|y_{ba_0}|$ as follows
{}
$$a^\alpha_{(0)}=-\eta^{\alpha\mu}\Big({\partial U_a\over\partial y^\mu_a}+
\sum_{b\not=a} \Big[{\partial U_b\over \partial y^\mu_a} -
 \Big<{\partial U_b\over \partial y^\mu_a}\Big>_a\Big] \Big)+\delta_a 
a^\alpha_{(0)}+$$
{}
$$+ \sum_{b\not=a}
\Bigg((4\beta-3\gamma-1){m_am_b\over y_{ba_0}}{n^\alpha\over y^2}
+2(\beta-1){m_am_b\over y}{N^\mu_{ba_0} \over {y^2_{ba_0}}}
\big(\delta^ \alpha_\mu+n^\alpha n_\mu\big)+$$
{}
$$+{m_am_b\over y^3_{ba_0}}{\cal P}_{\epsilon\lambda} 
\Big[(2\beta+{5\over3}\gamma)\eta^{\alpha\epsilon}n^\lambda+
(\beta-{1\over6})n^\alpha n^\epsilon n^\lambda\Big]\Bigg)+
{\cal O}(|y^\nu_a|)+{\cal O}(c^{-6}), 
\eqno(33)$$
\noindent where ${\cal P}^{\alpha\lambda}= \eta^{ \alpha \lambda}+
3N^\alpha_{ba_0}N^\lambda_{ba_0}$ is the polarizing operator, 
subscript $(a)$   denotes the planet Mercury and   the  post-Newtonian 
acceleration $\delta_a a^\alpha_{(0)}$ is due to the gravitational field of 
Mercury  only. This term  is  not new \cite{Den90} 
and it is  given by the expression $(C5a)$. 
Note that many relativistic terms in the geodetic equations have canceled out 
and, as a result, the equations of motion of a spacecraft around Mercury 
takes a very simple form Eq.(33).  At this point we have all the necessary 
equations in order to discuss  
gravitational experiments with the future {\it Mercury Orbiter } mission.

\section{Gravitational Experiment for Post 2000 Missions}
 
Mercury is the closest to the Sun of all the planets of the 
terrestrial group  and because of its
 unique location and orbital parameters, it is 
well suited to relativistic 
gravitational experiments. The short period of its solar orbit
 allows experiments over several orbital revolutions and its high 
eccentricity and inclination allow various effects to be well separated.
 In this Section  we will   discuss the possible gravitational
experiments  for the {\it Mercury Orbiter} mission.
Analysis  performed in this Section  is directed towards the future mission, 
so we will show which relativistic effects may be measured  and how 
accurately.

By means of a topographic Legendre expansion complete through the
second degree and order, the systematic error in Mercury radar
ranging has been reduced significantly \cite{And96a}.
However, a {\it Mercury Orbiter} is required before significant
improvements in relativity tests become possible.  Currently, the
precession rate of Mercury's perihelion, in excess of the 530
arcsec per century ($''$/cy) from planetary perturbations, is 
43.13 $''$/cy with a realistic standard error of 0.14  $''$/cy
 \cite{And91}.  After taking into account a small
excess precession from  solar oblateness, the later authors 
find that this result is consistent with general 
relativity.  Pitjeva
\cite{Pit93} has obtained a similar result but with a smaller estimated
error of 0.052  $''$/cy.  Similarly, attempts to detect a
time variation in the gravitational constant $G$ using Mercury's
orbital motion have been unsuccessful, again consistent with
general relativity.  The current result \cite{Pit93} is
$\dot{G}/G = (4.7 \pm 4.7) \times 10^{-12} \hbox{ yr}^{-1}$.  

Metric tests utilizing a  {\it  Mercury
Orbiter}  have been studied both at JPL and at the Joint Institute
for Laboratory Astrophysics (JILA) and the University of
Colorado.  The JPL studies, conducted in the 1970's, assumed that
orbiter tracking could provide daily measurements (normal points)
between the Earth and Mercury centers of mass with a 10~m
standard error.  A covariance analysis was performed utilizing a
16-parameter model consisting of six orbital elements for Mercury
and Earth respectively, the relativistic gravity parameters $\beta$
and $\gamma$, the solar quadrupole moment $J_{2\odot}$, and the
conversion factor AU between unit distance (Astronomical Unit) and
the distance in meters between Earth and Mercury.  It was assumed
that no other systematic effects were present, and that the
normal-point residuals after removal of the 16-parameter model
would be white and Gaussian.  The total data interval, assumed
equal to two years, corresponded to 730 measurements.  Under the
assumed random distribution of data, the error on the mean
Earth-Mercury distance was $10/\sqrt{730} = 37$ cm.  The JPL
studies showed, based on a covariance analysis, that the primary
metric relativity result from a {\it Mercury Orbiter} mission would be
the determination of the parameter $\gamma$, which 
describes the amount of spatial curvature caused by solar
gravitation.  The standard error was 0.0006, about a factor of
two improvement over the Viking Lander determination.  This
accuracy reflected the effect of spatial curvature on the
propagation of the ranging signal and also its effect on
Mercury's orbit, in particular the precession of the perihelion.
The error in the metric parameter $\beta$ and the error in the
solar $J_{2\odot}$ were competitive with current results, but not
significantly better.  

Within the last five years, a more detailed covariance analysis
by the JILA group \cite{Ash95} assumed 6~cm ranging
accuracy over a data interval of two years, but with only 40
independent measurements of range.  Unmodeled systematic errors
were accounted for with a modified worst-case error analysis.
Even so, the JILA group concluded that a two-order of magnitude
improvement was possible in the perihelion advance, the
relativistic time delay, and a possible time variation in the
gravitational constant {\small G} as measured in atomic units.  However,
the particular orbit proposed by ESA for its 2000 Plus mission was
not analyzed.  It is almost certain that the potential of the ESA
mission lies somewhere between the rather pessimistic JPL error
analysis and the JILA analysis of an orbiter mission more nearly
optimized for relativity testing.  

In order to study the relativistic effects in the motion of the 
{\it Mercury Orbiter} satellite, we  separate these effects into
the three following groups: 
\begin{itemize}
\item[(i).] The effects due to Mercury's motion with 
respect to the solar system barycentric reference frame.

\item[(ii).] Effects  in the satellite's motion with respect to the 
Hermean-centric reference frame.

\item[(iii).] Effects due to the dragging of the inertial frames.  
\end{itemize}
  
The effects of the first group 
are standard and all of them may be obtained directly from the 
Lagrangian function Eq.(6) or from the equations of motion Eqs.$(B1$-$B7)$.
 The effects of the second group can be discussed 
based on the equations (33). And finally, the effects of the last group 
can be discussed based on the coordinate transformation rules 
given by Eq.(22).  In the last case, however, we employ
a simplified version of these transformations, due to the limited expected 
accuracy ($\sim 1$ m) of the Mercury ranging data. Thus, in the future 
discussion we will use the following 
expression for the temporal components:
{} 
$$x^0(y^0_a, y^\mu_a)=y^0_a+c^{-2}\Bigg(\int^{y^0_a}\Big[\sum_{b\not=a} 
{m_b\over y_{ba_0}}
\Big(1+(I^{\lambda\mu}_a+I^{\lambda\mu}_b){{N_{ba_0}}_\lambda
{N_{ba_0}}_\mu\over y^2_{ba_0}}\Big)-$$
{}
$$-{1\over2}{v_{a_0}}_\mu v^\mu_{a_0}\Big]dt'-{v_{a_0}}_\mu y^\mu_a\Bigg)
+{\cal O}(c^{-4}),\eqno(34a)$$
\noindent where $I^{\mu\nu}_c$ represents the   intrinsic quadrupole 
moments Eq.(7) of the bodies. The corresponding expression for the spatial 
components of the coordinate transformation is given by: 
{}
$$x^\alpha(y^0_a, y^\mu_a)=y^\alpha_{a_0}(y^0_a)+y^\alpha_a+
c^{-2}\Bigg( {y_a}_\mu\Big[ \int^{y^0_a}\Omega_a^{\alpha\mu}(t') dt'-
{1\over 2} v^\alpha_{a_0} v^\mu_{a_0}-
\gamma \eta^{\alpha\mu}\sum_{b\not=a}{m_b\over y_{ba_0}}\Big]-$$
{}
$$-\sum_{b\not=a}{m_b\over y^2_{ba_0}}\Big[y^\alpha_a {y_a}_\mu N^\mu_{ba_0}  -
{1\over2}{y_a}_\mu y^\mu_a N^\alpha_{ba_0}\Big]+w^\alpha_a(y^0_a)\Bigg)+
{\cal O}(|y_a|^3)+{\cal O}(c^{-4}), \eqno(34b)$$
\noindent with the precession angular velocity tensor  $\Omega_a^{\alpha\beta}$ 
given as follows:
{}
$$\Omega_a^{\alpha\beta}(y^0_a)=\sum_{b\not=a}
\Big[(\gamma+{1\over2}){m_b\over y^2_{ba_0}}N^{[\alpha}_{ba_0}v^{\beta]}_{a_0}-
(\gamma+1){m_b\over y^2_{ba_0}}N^{[\alpha}_{ba_0}v^{\beta]}_{b_0}+$$
{}
$$ +(\gamma+1){m_b\over 2y^3_{ba_0}}{\cal P}^{[\alpha}_{\lambda}
(S^{\beta]\lambda}_a+S^{\beta]\lambda}_b)\Big],
\eqno(34c) $$
where $S^{\mu\nu}_c$ is the   intrinsic spin moments Eq.(8) of the bodies.

\subsection{Mercury's Perihelion Advance.}

Based on Mercury's barycentric equations of motion  
one can study the  phenomenon of Mercury's perihelion advance.
The secular trend in Mercury's 
perihelion\footnote{ 
It should be noted that  the {\it Mercury Orbiter} itself, 
being placed in orbit around Mercury,   will experience the 
phenomenon of periapse advance as well. 
However, we expect that uncertainties in Mercury's gravity field will mask the 
relativistic precession, at least at the  level of interest for 
ruling out alternative gravitational theories.}
 depends on the linear combination of the  {\small PPN} 
parameters $\gamma$ and $\beta$ and the solar quadrupole 
coefficient $J_{2\odot}$ (\cite{Wil93}, \cite{Nob86},
\cite{Hei90}):
{}
$$\dot{\pi} =  (2+2\gamma-\beta){  \mu_\odot n_M \over  a_M(1-e^2_M)} + 
 {3\over 4}\Big({R_\odot \over a_M}\Big)^2 
{J_{2\odot}n_M\over (1-e^2_M)^2 } (3\cos^2 i_M-1), 
\hskip 10pt ''/\hbox{cy} \eqno(35a)$$
\noindent where $a_M, n_M, i_M$ and $e_M$ are the 
mean distance, mean motion, inclination and eccentricity of Mercury's orbit.
The parameters $\mu_\odot$ and $R_\odot$ are the solar gravitational constant  
and  radius respectively. 
For  Mercury's orbital parameters  one obtains:
{}
$$\dot{\pi}  = 42 {''\hskip -3pt .}98 \hskip 2pt
\Big[ \hskip 1pt{1\over 3}(2+2\gamma-\beta) +0.296 
\cdot {J_{2\odot}}\times 10^4\Big], \hskip 10pt ''/\hbox{cy}  \eqno(35b)$$
 
\noindent Thus, the accuracy of the relativity tests on the {\it Mercury Orbiter}
 mission will depend on   our knowledge of the solar gravity field.
The major source of  uncertainty in these measurements is 
the solar quadrupole  moment $ {J_2}_\odot$. 
  As evidenced by the oblateness of the photosphere 
\cite{Bro89} and perturbations in frequencies of solar
oscillations, the internal structure of the Sun is slightly
aspherical.  The amount of this asphericity is uncertain. 
It has been suggested that it could be significantly larger than
calculated for a simply rotating star, and that the internal 
rotation rate varies with the
solar cycle \cite{Goo91}.  Solar oscillation data
suggest that most of the Sun rotates slightly slower than the
surface except possibly for a more rapidly rotating core \cite{Duv84}.
  An independent measurement  of $ {J_2}_\odot$
performed  with the {\it Mercury Orbiter} 
would provide a valuable direct confirmation of the indirect
helioseismology value $(2\pm 0.2)\times 10^{-7}$.  
Furthermore, there are suggestions of a
rapidly rotating core, but helioseismology determinations are 
limited by uncertainties at depths below 0.4 solar radii 
\cite{Lib91}.  

The {\it Mercury Orbiter} will help us understand  this
asphericity and  independently will enable us to gain some     
important data on the properties of the solar interior and the 
features of it's rotational motion.  Preliminary 
analysis of a {\it Mercury Orbiter} mission suggests 
that $J_{2\odot}$ would  be measurable to at best
 $\sim 10^{-9}$  \cite{Ash95} or about 1\% of the expected  
$J_{2\odot}$ value. This should be compared with the  present $10\%$   
solar oscillation determination \cite{Bro89}.

 \subsection{The Redshift Experiment.}

Another important experiment that could be performed on a {\it Mercury
Orbiter} mission is a test of the solar gravitational redshift.
This would require  a stable frequency standard to be  flown on the spacecraft.
The experiment would provide a fundamental test of the theory of
general relativity and the Equivalence Principle upon which it
and other metric theories of gravity are based (\cite{Wil93}, \cite{Sha76}).
Because in general relativity the gravitational redshift of an
oscillator or clock depends upon its proximity to a massive body
(or more precisely the size of the Newtonian potential at its
location), a frequency standard at the location of Mercury would
experience a large, measurable redshift due to the Sun.
With  the result  for the   function
${\cal K}_a$ given by Eq.$(22a)$ and Eq.$(34a)$ in hand,  one can obtain  the 
corresponding Newtonian proper frequency variation  between the barycentric 
standard of time and that of the satellite  (the  terms with  the magnitude 
up to $10^{-12}$), given as:
{} 
$${dx^0\over dy^0_{(0)}}=1+ {\mu_\odot\over R_M} +{\mu_M\over y_{(0)}}+
{1\over2c^2}({\vec v}_M  + {\vec v}_{(0)} )^2-{\mu_\odot\over R^3_M}
({\vec R}_M {\vec y}_{(0)})  +{\cal O}(c^{-4}), \eqno(36a)$$
\noindent where $(y^0_{(0)},{\vec y}_{(0)})$ are the 
four-coordinates of the spacecraft in the Hermean-centric reference frame 
and ${\vec v}_{(0)}$ is the spacecraft orbital   velocity. 
One can see that  the eccentricity of Mercury's orbit would be highly effective 
in varying the solar potential at the clock, thereby producing a
distinguishing signature in the redshift. The anticipated frequency 
variation between perihelion and aphelion is to first-order in
the eccentricity:   
{}
$$\Big({\delta f\over f_0}\Big)_{e_M}=   
{2\mu_\odot e_M \over  a_M}.\eqno(36b)$$

\noindent This contribution  
is quite considerable and is calculated to be 
$(\delta f/f_0)_{e_M}=1.1\times 10^{-8}$. 
Its magnitude, for instance, 
at a radio-wave length $\lambda_0=3$ cm ($f_0=10$ GHz) 
is $ (\delta f)_{e_M}=  110$ Hz.  
 We would also benefit
from the short orbital period of Mercury, which would permit the
redshift signature of the Sun to be measured several times over
the duration of the mission. 
If the spacecraft tracking and modelling are of sufficient precision to 
determine the 
spacecraft position relative to the sun to $100 \hskip 1pt \rm m$ (a conservative 
estimate)  then 
a frequency standard with $10^{-15}$ fractional frequency stability 
$\delta f/f=10^{-15}$ would be able to measure the redshift to 1 part in $10^7$ 
or better. This stability is within the capability of proposed spaceborne 
trapped-ion \cite{Pre92}
or H-maser clocks (\cite{Ves80}, \cite{Wal94}).
 
\subsection{The SEP violation effect.}

Besides the   Nordtvedt   effect 
(for more details see \cite{AndTur96} and \cite{And96}), there  exists an 
interesting possibility for testing the  {\small SEP}  violation effect  
by studying  spacecraft   motion in orbit around Mercury. The corresponding 
equation of motion is given by Eq.(33).   
As one can see, the  two terms in the second line of this equation   vanish
 for   general relativity, but  for scalar-tensor theories,
they   become responsible for  
small deviations of the spacecraft  motion from the fiducial  geodesic. 
Both of these effects, if they exist,  are due to non-linear coupling of the 
gravitational field of Mercury to external gravity. They come from the    
expression for  $W_a$ given by Eq.$(28a)$,
which is the local post-Newtonian contribution to the $g_{00}$ component  of 
the metric tensor in the proper reference frame.

The  first of these terms may be interpreted as a
dependence of the locally measured   gravitational constant 
on the  external gravitational environment and may be expressed in the 
vicinity of  body $(a)$ as follows:
{}  
$$G_a=G_0\Big[1-(4\beta-3\gamma-1)\sum_{b\not=a}{m_b\over y_{ba_0}} \Big]. 
\eqno(37)$$
\noindent  In the case of a satellite around Mercury, the main contribution to 
this effect  
 comes from the Sun\footnote{Note  that this combination of  {\small PPN}  
parameters  differs 
from that  for a similar effect presented in \cite{Wil93}. The reason for this  
is that, in our case the   transformations  in the form of Eq.(22) 
enable us  to define the physically adequate transformation rules 
for the   metric tensor between
the barycentric  and    planeto-centric  reference frames and, hence, to 
obtain the correct and complete equations of geodesic motion Eq.(33) in the  
Hermean-centric reference frame.}. Because of the high  eccentricity of 
Mercury's orbit, the periodic changing of the sun's {\it local} gravitational 
potential may produce an observable effect, which can be modeled by a  
periodic time variation in the effective {\it local} gravitational constant:
{}
$$ \Big[{\dot{G}\over  G}\Big]_{\rm period}=
(4\beta-3\gamma-1)\Big[{\mu_\odot\over a_M(1-e_M^2)}\Big]^{3\over2}
{c e_M\sin\phi(t)\over a_M(1-e_M^2)}\big(1+e_M\cos\phi(t)\big)^2,\eqno(38)$$
\noindent which gives the following estimate for this effect  on Mercury's orbit:
{}
$$\Big[{\dot{G}\over  G}\Big]_{\rm period}\approx(4\beta-3\gamma-1) 
\times 1.52\times 10^{-7}\sin\phi(t) \qquad   {\rm yr^{-1}}.\eqno(39)$$

Note that this  effect Eq.(39)    is fundamentally 
different from that introduced by Dirac's  hypothesis of possible  
time dependence  of the gravitational constant \cite{Pit93}.
As one can see from expression (39), the characteristic time   
in this case is Mercury's siderial period.
This short period may be considered as an advantage from the experimental 
point of view. In addition,   the results of the redshift experiment could help 
in confident studies of this effect.
Recently a different combination of the post-Newtonian parameters 
in the Nordtvedt effect, $\eta=4\beta-\gamma-3$, was  measured at  
$\eta\le 10^{-4}$ \cite{Dic94}. This means  that, in order to 
obtain the  comparable accuracy for the combination of parameters Eq.(39), 
one should perform the   Mercury  gravimetric measurements  
 on the level no less precise than $\big[{\dot{G}/ G}\big]_{\rm period} 
\approx 10^{-11} \hskip 3pt  {\rm yr}^{-1}$. 
Recently a   group at the University of Colorado has analyzed 
a  number of gravitational experiments possible  with  future 
  Mercury   missions \cite{Ash95}.
Using a modified worst case error analysis, this group suggests that  after 
one year  of ranging between Earth and Mercury (and assuming a 6 cm rms error), 
the fractional accuracy of determination of the sun's gravitational constant 
$m_\odot G$ is expected to be of order $\sim 2.1 \times 10^{-11}$. 
Moreover, even higher accuracy could be achieved  with a 
Mercury lander   as proposed by  
\cite{Ash95}. This suggests that the experiment 
for determination of the effect Eq.(39) may be feasible 
with the {\it Mercury Orbiter} mission. 
      
Another interesting effect on the satellite's orbit may be derived
from the Eq.(33) in the form of following acceleration term: 
{}
$$\delta {\vec a}_{(0)SEP}= 2(\beta-1){m_Mm_\odot\over yR_M^2} 
\big({\vec N_M}- {\vec n} ({\vec n}{\vec N_M})\big), \eqno(40)$$
\noindent where  $R_M$ is Mercury's heliocentric radius-vector
 and $\vec{N}_M$ is the unit vector along this direction. 
This effect is very small for the  orbit proposed for ESA's {\it Mercury Orbiter} 
mission. However, one can show that there exist two resonant orbits for a 
satellite around Mercury, either with the orbital frequency $\omega_{(0)}$ equal 
to Mercury's siderial frequency $\omega_M$:  $\omega_{(0)}\approx\omega_M$ or at 
one third of this frequency $\omega_{(0)}\approx \omega_M/3$.
For these resonant orbits,  the corresponding experiment 
could provide  an independent direct test of the  parameter $\beta$.

\subsection{The Precession Phenomena.}

In addition to the perihelion advance, while constructing the 
Hermean proper reference frame,  one should take into account  several 
precession phenomena included in the transformation function ${\cal Q}^\alpha_a$
and associated with the angular momentum of the bodies. 
As one may see directly from Eq.$(22b)$ and Eq.$(34b)$,  besides the obvious  
special relativistic contributions, the    post-Newtonian 
transformation of the spatial coordinates 
contains   terms due to the non-perturbative  influence of the gravitational 
field. This non-Lotentzian behavior of 
the post-Newtonian transformations was discussed first by \cite{Chandra67} 
for the case of post-Galilean transformations. Our derivations 
differ from the latter by taking into account the acceleration of the proper 
reference frame  and by including the infinitesimal precession of the coordinate 
axes with the angular velocity tensor $\Omega^{\alpha\beta}_M$, given by 
expression $(34c)$ as follows: 

$$\Omega_M^{\alpha\beta} =\sum_{b\not=M}
\Big[(\gamma+{1\over2}){m_b\over y^2_{bM_0}}N^{[\alpha}_{bM_0}v^{\beta]}_{M_0}-
(\gamma+1){m_b\over y^2_{bM_0}}N^{[\alpha}_{bM_0}v^{\beta]}_{b_0}+ $$
{}
$$+(\gamma+1){m_b\over 2y^3_{bM_0}} {\cal P}^{[\alpha}_{\lambda}
(S^{\beta]\lambda}_M+S^{\beta]\lambda}_b)\Big].\eqno(41) $$

\noindent where the summation is performed 
over the other bodies of the solar system. 
This expression re-derives and generalizes the result for the precession of  
the spin of a gyroscope  $\vec{s_0}$ attached to a test body orbiting 
a gravitating primary. Previously this result  was obtained from the  theory of  
Fermi-Walker transport \cite{Wil93}.  Indeed, in accord  with Eq.$(34b)$, 
this spin (or coordinate axes of a proper Hermean reference frame)
will precess  with respect to a distant standard of 
rest such as quasars or distant galaxies. The   motion of the spin vector of a 
gyroscope  can be described by the relation: 
{}
$${d  \vec{s_0}\over dt}=[ \vec{\Omega}_M \times  \vec{s_0}].\eqno(42)$$

\noindent By keeping   the leading contributions only and neglecting the 
influence of the Mercury's intrinsic spin moment, we obtain from the expression 
(41)   the angular velocity $\vec{\Omega}_M$ in the  following form:
{}
$${\vec \Omega}_M= (\gamma+{1\over2}){\mu_\odot\over R_M^3}[{\vec R}_M\times 
{\vec v}_{M_0}]-
(\gamma+1){\mu_\odot\over R^3_M}[{\vec R}_M\times {\vec v}_\odot]+ $$
{}
$$+(\gamma+1){\mu_\odot\over 2R^3_M}\Big(\vec{S_\odot}-
{3(\vec{S_\odot}\vec{N}_M)\vec{N}_M}\Big), \eqno(43)$$

\noindent where $\vec{v}_{M_0}$  and $\vec{v}_\odot$  are Mercury's and the Sun's
 barycentric orbital velocities and 
${\vec S}_\odot$ is the solar intrinsic spin moment. 

The first term in Eq.(43) is known as  geodetic precession \cite{DeS16}. 
This term arises in any non-homogeneous gravitational field  because of the 
parallel transport of a direction defined by $\vec{s_0}$. It can be 
viewed as spin precession caused by a  coupling 
between the particle velocity $\vec{v}_{M_0}$ and the static  
part of the space-time geometry.
For  Mercury orbiting the Sun this precession has the form:
{}  
$$\vec{\Omega}_G= (\gamma +{1\over 2})
{\mu_\odot\over R^3_M}(  \vec{R}_M\times \vec{ v}_{M_0}). \eqno(44)$$
\noindent 
This effect could be studied for the {\it Mercury Orbiter}, which, being 
placed in orbit around  Mercury is in effect a gyroscope orbiting the Sun.
Thus, if we    introduce the angular momentum per unit mass, 
$ \vec{ L}= \vec{R}_M\times  \vec{v_{M_0}}$, 
of   Mercury in solar orbit, the equation (44) shows
that $ \vec{ \Omega}_G$ is directed along the pole of the ecliptic, 
in the direction of $ \vec{ L}$. The vector $ \vec{\Omega}_G$ has a constant part
{}
$$ \vec{\Omega}_0={1\over 2}(1+2\gamma) 
 {\mu_\odot\omega_M\over a_M}=  {1+2\gamma\over 3}\cdot 0.205  
 \hskip 10pt  { ''/\hbox{yr}}, \eqno(45a) $$

\noindent with a significant correction   due to the eccentricity $e_M$
of the Mercury's orbit,
{}
$$ \vec{\Omega}_1 \cos \omega_M t=
 {3\over 2}(1+2\gamma) {\mu_\odot\omega_M\over a_M} e_M\cos \omega_M t_0= 
{1+2\gamma\over 3}\cdot0.126 \cos \omega_M t_0
 \hskip 10pt  {''/\hbox{yr}}, \eqno(45b) $$

\noindent where $\omega_M$ is   Mercury's siderial frequency,   
$t_0$ is reckoned from a perihelion passage; 
$a_M$ is the semimajor axis of  Mercury's orbit. 

Geodetic precession has been studied for the motion of lunar perigee 
and its existence   was first confirmed with
an accuracy of 10\% \cite{Ber87}.
Two other groups have analyzed the lunar laser-ranging data more
completely to estimate the deviation of the lunar orbit from the  
predictions of general relativity  
 (\cite{Sha88}, \cite{Dic89}). 
Geodetic precession has been confirmed  within a standard deviation of 2\%.  
The precession of the  orbital plane proposed for ESA's {\it Mercury orbiter}
(periherm at 400 km altitude, apherm at 16,800 km, period 13.45 hr and latitude
of  periherm at +30 deg) would   include a contribution of 
order 0.205 ${''/\hbox{yr}}$  from the geodetic precession.
We recommend that this
precession   be included in future studies of the 
{\it Mercury Orbiter} mission.

The third term in   expression (43) is known as 
Lense-Thirring precession $\vec{\Omega}_{LT}$. 
This term gives the 
relativistic precession of the gyroscope's spin 
$\vec{s}_0$ caused by the intrinsic angular momentum 
$\vec{S}$ of the central body.
This effect is responsible for a small perturbation in the  orbits 
of artificial satellites around the Earth (\cite{Tap72},
\cite{Rie91}).
However,  our preliminary studies indicate that 
this effect is so small for the satellite's orbit around Mercury 
that will be masked by uncertainties in the orbit's inclination.

\section{Discussion}
 
The  use of tracking data  from orbiters for relativity tests requires 
the consideration of more error sources than for landers. 
This is because of the need to convert from the measured earth-spacecraft 
distance to the desired earth-planet distance. This involves determining the 
orbit of the spacecraft about the 
planetary center of mass, which requires solving from the tracking data  
for a number of spatial harmonics of the gravitational field, solving 
for radiation pressure, and other non-gravitational forces.
Firing  of attitude control
jets which produce unbalanced forces are of particular concern. The 
orbit determination of the Mars orbiter {\it Mariner 9}, for example,  was  
substantially affected not only   by these factors, but also by  the fact that 
the spacecraft was placed on a 12 hr period orbit with low periapsis 
(\cite{And78}, \cite{And86}). Thus, in order to precisely describe the motion 
of the {\it Mercury Orbiter}  relative to Earth, one must solve two problems, 
namely: (i) the problem of the satellite motion about Mercury's center of mass 
in the Hermean-centric frame, and (ii) the relative motion of the 
both planets - Earth and Mercury - in the solar system barycentric reference frame.
Our analysis provides a framework for the complete solution of these two 
problems in terms of  the   corresponding differential equations.  

The formalism  presented in  this paper  addresses the general problem of radio
tracking of a {\it Mercury Orbiter}. We have presented the Hermean-centric 
equations of the satellite motion, the barycentric equations of the planet's 
motion in the solar system barycentric reference frame, and the coordinate 
transformations which   link  these equations together. In particular, our 
analysis shows that in a  proper Hermean-centric reference frame, the 
equations of the satellite motion  depend on   Mercury's gravitational field only. 
 This set of equations is available in the form of the  
motion of test bodies  in the isolated gravitational 
one body problem. 
  The existence of the external gravitational field  manifests
itself in the form of the usual tidal forces, but it also determines   
the dynamic properties of the constructed 
 Hermean-centric proper reference frame.  
Within the accuracy expected for the future {\it Mercury Orbiter} mission,
one can  completely neglect the post-Newtonian tidal terms. 
However,   while constructing this  reference frame,
we derived terms smaller than the expected accuracy of future experiments.
Indeed, the last term in  equation (33) is due to the coordinate 
transformation to the Fermi-normal-like  reference frame  in the 
planet's vicinity. One can neglect this term for solar system motion. However, 
if one  applies the results to  problems of motion within a more intensive 
gravitational environment,  this term can play  a significant role.
The  application of the results obtained here to problems of  motion of   double 
pulsars is currently under study and will be reported later.

\vskip 18pt\noindent
{\Large\bf Acknowledgements}
\vskip 11pt\noindent
SGT acknowledges the  support  by the  National Research Council
of the USA.
The research described in this paper was carried out by the Jet
Propulsion Laboratory, California Institute of Technology, and
was sponsored by the Ultraviolet, Visible, and Gravitational
Astrophysics Research and Analysis Program through an agreement
with the National Aeronautics and Space Administration.

\vskip 18pt\noindent 
{\large\bf Appendix A:  Generalized Gravitational Potentials.}
\vskip 11pt\noindent 
 The generalized gravitational potentials for the non-radiative 
problems in the  {\small WFSMA} are given as in   \cite{Wil93}:
{}
$$U(z^p)  = \int { d^3z' \rho  (z'^p)\over |z^\nu - z'^\nu|}, 
\qquad V^\alpha(z^p)  = - \int d^3z' {\rho (z'^p)v^\alpha (z'^p)\over 
|z^\nu-z'^\nu|},$$
{}
$$W^\alpha(z^p)  = \int d^3z' \rho (z'^p)v_\mu(z'^p)
{(z^\alpha-z'^\alpha)(z^\mu-z'^\mu)\over|z^\nu-z'^\nu|},$$
{}
$$A(z^p)  = \int d^3z' \rho (z'^p){ [v^\mu(z'^p)(z^\mu-z'^\mu)]^2
\over|z^\nu-z'^\nu|^3},
\qquad \chi(z^p)  = -\int  d^3z'  \rho  (z'^p)  |z^\nu - z'^\nu|, $$ 
{}
$$U^{\alpha\beta}(z^p)  = \int d^3z' \rho (z'^p){(z^\alpha-z'^\alpha)
(z^\beta-z'^\beta)\over|z^\nu-z'^\nu|^3},$$
{}
$$\Psi(z^p)= -(\gamma+1)\Phi_1 -  (3\gamma+1-2\beta)\Phi_2-
 \Phi_3 -3\gamma\hskip 1pt\Phi_4, $$

\noindent where the other potentials are given as  follows:
{}
$$\Phi_1(z^p)  = - \int d^3z'  {\rho (z'^p) v_\lambda 
(z'^p)v^\lambda(z'^p)\over 
|z^\nu  - z'^\nu|},\qquad \Phi_{2} (z^p) 
= \int d^3z' {\rho (z'^p)U(z'^p)\over |z^\nu - z'^\nu |}, $$
{}
$$\Phi_{3} (z^p) =  \int  d^3z' { \rho (z'^p)\Pi(z'^p)\over 
|z^\nu  -z'^\nu|}, \qquad \Phi_{4} (z^p) = 
\int d^3z' {\rho (z'^p)  p(\rho(z'^p))\over |z^\nu- z'^\nu|},$$
{}
$$\Phi_w(z^p)  = \int\hskip -3pt\int d^3z'  d^3z''\rho (z'^p)\rho (z''^p)
{(z^\beta-z'^\beta)\over|z^\nu-z'^\nu|^3} 
\Big[{(z_\beta-z''^\beta)\over|z'^\nu-z''^\nu|}
-{(z'^\beta-z''^\beta)\over|z^\nu-z''^\nu|}
\Big]. $$

\noindent In order to indicate the functional dependence in the potentials 
introduced above, we have  used the following notation: 
$(z^p)\equiv (z^0, z^\nu)$. Then for any function $f$
one will have: $f(z^p)=f(z^0,z^\nu)$ and $f(z'^p)=f(z^0,z'^\nu)$.

\vskip 18pt\noindent 
{\large\bf Appendix B:   Simplified Barycentric Equations of Motion.}
\vskip 11pt\noindent
In this Appendix we will present the simplified barycentric equations of motion
corresponding the Lagrangian Eq.(6). Assuming that bodies in the system possess 
the lowest intrinsic multipole moments only,
one can obtain the corresponding simplified equations of motion. Thus, with the 
help of the expressions  (6), for an arbitrary  body $(a)$ these equations will 
read as follows: 
{}
$$\ddot{r}^\alpha_a =\sum_{b\not=a} 
{{\cal M}_b \over r_{ab}^2}{\widehat{n}_{ab}}^\alpha +
\sum_{b\not=a} {m_b\over r_{ab}^2}\Big[{\cal A}^\alpha_{ab}+
{{{\cal B}^\alpha_{ab}}\over r_{ab}}+ {{\cal C}^\alpha_{ab} \over r_{ab}^2}-$$
{}
$$-{n_{ab}^\alpha\over r_{ab}}\Big((2\beta+2\gamma-
2\tau+1)m_a+(2\beta+2\gamma-2\tau)m_b\Big)\Big]+$$
{}
$$+ \sum_{b\not=a}\sum_{c\not=a,b}m_bm_c 
{\cal D}^\alpha_{abc}+{\cal O}(c^{-6}), \eqno(B1)$$

\noindent where, in order to account for the influence of the  
gravitational binding energy $E_b$, 
we have introduced the passive gravitational rest mass ${\cal M}_b$ 
(\cite{Nor68b}, \cite{Wil93})  as follows 
{}
$${\cal M}_b = m_b\Big(1+(3+\gamma-4\beta)E_b+{\cal O}(c^{-4})\Big). \eqno(B2)$$

\noindent The unit vector  $n_{ab}$ must also be   
corrected using the gravitational binding energy
and the  tensor of the quadrupole moment $I^{\alpha\beta}_a$ of the body $(a)$
under question:
{}
$$\widehat{n}^{\alpha}_{ab} = n^{\alpha}_{ab}\Big(1+(3+\gamma-4\beta)E_a+
5{n_{ab}}_\lambda {n_{ab}}_\mu {I^{\lambda\mu}_a\over r_{ab}^2}\Big)+
2{n_{ab}}_\beta {I^{\alpha\beta}_a\over r_{ab}^2}+{\cal O}(c^{-4}). 
\eqno(B3)$$
 
\noindent The term ${\cal A}^\alpha_{ab}$ in the expression (B1) 
is the orbital term  which is given as follows:
{}
$${\cal A}^\alpha_{ab}= v^\alpha_{ab} {n_{ab}}_\lambda\Big(v^\lambda_a-
(2\gamma-2\tau+1)  v^\lambda_{ab}\Big)+$$
{}
$$+ n^\alpha_{ab}\Big({v_a}_\lambda v^\lambda_a-
(\gamma+1+\tau) {v_{ab}}_\lambda v^\lambda_{ab}-
3\tau ({n_{ab}}_\lambda {v_{ab}}^\lambda)^2-
{3\over 2}({n_{ab}}_\lambda v^\beta_{b})^2\Big). \eqno(B4)$$

\noindent The spin-orbital term $ {\cal B}^\alpha_{ab} $ has the form: 
{}
$$ {\cal B}^\alpha_{ab} =({3\over 2}+2\gamma){v_{ab}}_\lambda
(S^{\alpha\lambda}_a+S^{\alpha\lambda}_b)+{1\over2}{v_a}_\lambda
(S^{\alpha\lambda}_a-S^{\alpha\lambda}_b)+$$
{}
$$ +{3\over 2}(1+2\gamma){n_{ab}}_\lambda{v_{ab}}_\beta\Big[n_{ab}^\beta 
(S^{\alpha\lambda}_a+S^{\alpha\lambda}_b)-{n_{ab}}^\alpha 
(S^{\beta\lambda}_a+S^{\beta\lambda}_b)\Big]+$$
{}
$$+{3\over 2}{n_{ab}}_\lambda \Big[n_{ab}^\alpha 
({v_{a}}_\beta S^{\beta\lambda}_b-{v_{b}}_\beta S^{\beta\lambda}_a) + 
{n_{ab}}_\beta{v_{ab}}^\beta S^{\alpha\lambda}_b\Big]. \eqno(B5)$$

\noindent  The term ${\cal C}^\alpha_{ab}$ is 
caused by the oblateness of the bodies in the system: 
{}
$$ {\cal C}^\alpha_{ab} =2{n_{ab}}_\beta I^{\alpha\beta}_b +
5n_{ab}^\alpha{n_{ab}}_\lambda {n_{ab}}_\mu I^{\lambda\mu}_b.
 \eqno(B6) $$

\noindent And, finally, the contribution ${\cal D}^\alpha_{abc}$ 
to the equations of motion Eq.$(B1)$ of  body $(a)$, caused by the 
interaction of the other planets ($b\not=a, c\not=a,b$) 
 with each other is presented as:
{}
$${\cal D}^\alpha_{abc}={n^\alpha_{ab}\over 
r_{ab}^2}\Big[(1-2\beta){1\over r_{bc}}- 
2(\beta+\gamma){1\over r_{ac}}\Big]+$$
{}
$$+\tau{\Pi^{\alpha\lambda}_{ab}
\over r^3_{ab}}({n_{bc}}_\lambda+{n_{ca}}_\lambda)  
+\tau{{n_{ab}}_\lambda\over r_{ab}^2}
{\Lambda^{\alpha\lambda}_{ac}\over r_{ac}}+
{1\over 2}(1+2\tau){{n_{bc}}_\lambda\over r_{bc}^2}
{\Lambda^{\alpha\lambda}_{ac}\over r_{ac}}+
2(1+\gamma){{n_{bc}}^\alpha\over r_{bc}^2r_{ab}}, \eqno(B7)$$

\noindent where
$\Lambda_{ab}^{\mu\nu}=\eta^{\mu\nu}+n_{ab}^\mu n_{ab}^\nu$  
and $\Pi_{ab}^{\mu\nu}=\eta^{\mu\nu}+3n_{ab}^\mu n_{ab}^\nu$.

The metric tensor Eq.(3), the Lagrangian function Eq.(6) and the equations 
of motion Eqs.$(B1$-$B7)$  define  the  
behavior  of the celestial  bodies in the post-Newtonian 
approximation. These equations may be considerably simplified  by
taking into account that 
the leading contribution to these equations is the 
solar gravitational field. Hence  they may be used for
producing the numerical codes in relativistic orbit determination 
formalisms for planets and satellites (\cite{Stand92}, \cite{Moy81}, 
\cite{Rie91}, \cite{Hua90}) as well as for
analyzing   the gravitational experiments in the solar system
(\cite{Wil93}, \cite{Pit93}, \cite{And96}).
Moreover, taking into account the expected accuracy of the orbit 
determination   for the future {\it Mercury Orbiter} mission,
one may neglect the post-Newtonian perturbations caused by the other planets 
on the Mercury's orbital motion in the terms ${\cal A}^\alpha_{ab}, 
{\cal B}^\alpha_{ab}, {\cal C}^\alpha_{ab},
{\cal D}^\alpha_{ab}$, leaving only the solar contribution.
 The   set of the resulting equations now defines  the properties of the   
 motion of the  Mercury in the solar system barycentric reference frame.

\vskip 18pt\noindent
{\large\bf Appendix C:  Equations of the Spacecraft Motion.}
\vskip 11pt\noindent 
In this Appendix we will study the motion of the spacecraft around the planet 
Mercury. 
In order to obtain the Hermean-centric equations of the satellite motion we will
write out the equations of geodesics to the required degree of accuracy.
For $n=\alpha$ we have:
{}
$${du^\alpha\over ds}+\Gamma^\alpha_{00}u^0u^0+
2\Gamma^\alpha_{0\beta} u^0u^\beta+\Gamma^\alpha_{\mu\beta}u^\mu u^\beta=
{\cal O}(c^{-6}). \eqno(C1) $$
\noindent We consider the metric tensor of Riemann  space-time  
to be  given by the expressions (30) in this case.
It is then possible to find the connection components of this  space-time
needed for subsequent computations:
{}
$$\Gamma^{\alpha}_{00}(y^p_{a}) = \eta^{\alpha\lambda}\Big[ 
{\partial\overline{U}\over \partial y^\lambda_a}
- {\partial W_a\over \partial y^\lambda_a} 
-\gamma   {\partial U^2_a\over \partial y^\lambda_a}\Big] +
2(\gamma+1){\partial V^\alpha_a\over \partial y^0_a}+$$ 
{}
$$ + \Bigg((2\gamma-1) a^\alpha_{a_0} {a_{a_0}}_\mu - 
\gamma\delta^\alpha_\mu \cdot a^\lambda_{a_0} {a_{a_0}}_\lambda -
\eta^{\alpha\lambda}\sum_{b\not=a}\Big[
\Big<\partial^2_{\lambda\mu} W_b\Big>_a+
 2\gamma U_a \Big<\partial^2_{\lambda\mu} U_b\Big>_a\Big]+$$
{}
$$ +\sum_{b\not=a}{\partial \over \partial y^0_a}
\Big[(\gamma+1)  \Big<\partial_{(\mu} V^{\alpha)}_b \Big>_a -
\delta^\alpha_\lambda{\partial \over \partial y^0_a}
\Big<U_b\Big>_a\Big]\Bigg) \cdot y^\mu_a +
{\cal O}(|y^\nu_a|^2)+ {\cal O}(c^{-6}), \eqno(C2a)$$
{}
$$\Gamma^{\alpha}_{0\beta}(y^p_{a}) =\gamma\delta^\alpha_\beta 
{\partial U_a \over \partial y^0_a} + 
(\gamma+1)\delta^\mu_{[\beta}{\partial V^{\alpha]}_a  \over\partial y^{\mu}_a} + 
\gamma  \eta_{\beta\mu}{\dot a}^{[\mu}_{a_0}y^{\alpha]}_a +$$
{}
$$+(\gamma+1) \sum_{b\not=a}\Big[\Big<\partial^2_{\beta\mu} V^\alpha_b\Big>_a
-\Big<\partial^\alpha\partial_\mu {V_b}_\beta \Big>_a
\Big]\cdot y^\mu_a+{\cal O}(|y^\nu_a|^2)+ {\cal O}(c^{-5}), \eqno(C2b)$$
{}
$$\Gamma^{\alpha}_{\beta\omega}(y^p_{a}) = 
 \gamma\Big(\delta^\alpha_\beta {\partial U_a\over \partial y^\omega_a}  + 
\delta^\alpha_\omega {\partial U_a\over \partial
 y^\beta_a}  -\eta_{\beta\omega}\eta^{\alpha\lambda}
{\partial U_a\over \partial y^\lambda_a}\Big)+$$
{}
$$ +{1\over3}\gamma\sum_{b\not=a}\Big[\eta_{\beta\mu}
\Big<\partial^\alpha\partial_\omega U_b\Big>_a  +
\eta_{\omega\mu}\Big<\partial^\alpha\partial_\beta U_b\Big>_a+
\delta^\alpha_\omega\Big<\partial^2_{\beta\mu} U_b\Big>_a+
\delta^\alpha_\beta\Big<\partial^2_{\omega\mu} U_b\Big>_a-$$
{}
$$-2\eta_{\beta\omega}\Big<\partial^\alpha\partial_\mu U_b\Big>_a-
2\delta^\alpha_\mu\Big<\partial^2_{\beta\omega} U_b\Big>_a\Big]
\cdot y^\mu_a+{\cal O}(|y^\nu_a|^2)+ {\cal O}(c^{-4}).\eqno(C2c) $$

To reduce the equation of geodesic motion Eq.$(C1)$ we shall use 
these expressions above and the definition for the four-vector of velocity 
in the form:
{}
$$u^n={dy^n_a\over dy^0_a}\Big(g_{00}+2g_{0\mu}v^\mu+g_{\mu\lambda}v^\mu 
v^\lambda\Big)^{-1/2}.$$
\noindent Then  by taking into account  that $d/ds=u^0d/dy^0_a$ 
(with the components of the three-dimensional velocity
vector of the point body denoted as $v^\alpha_{(0)}=dy^\alpha_a/dy^0_a$) and  by
  using the Newtonian equation of motion of a point body as:
{}
$$a^\alpha_{(0)}= {d v^\alpha_{(0)}\over dy^0_a}= -\eta^{\alpha\mu}
{\partial {\overline U}\over\partial y^\mu_a}+{\cal O}(c^{-4}),$$
\noindent we may make the following simplification:
{}
$$v^\alpha_{(0)}{d\ln u^0\over dy^0_a}=v^\alpha_{(0)}\Big({\partial 
{\overline U}\over\partial y^0_a}+
2v^\mu_{(0)}{\partial {\overline U}\over\partial y^\mu_a}+{\cal O}(c^{-5})\Big).$$

\noindent Substituting this relation into the equations of motion $(C1)$, 
we find the acceleration $a^\alpha_{(0)}$ of the point body:
{}
$$a^\alpha_{(0)}=-\eta^{\alpha\mu}{\partial {\overline U}\over\partial y^\mu_a}
(1-2\gamma U_a)+  \partial^\alpha W_a +
\sum_{b\not=a}\Big<\partial^\alpha\partial_\mu W_b\Big>_a \cdot y^\mu_a- $$
{}
$$ -(2\gamma+1)v^\alpha_{(0)}{\partial U_a\over\partial y^0_a}-
2(\gamma+1){\partial   V^\alpha_a \over\partial y^0_a}-2(\gamma+1) 
{v_{(0)}}_\mu\Big[ \partial^\mu V^\alpha_a-\partial^\alpha V^\mu_a\Big] +$$
{}
$$+ \gamma {v_{(0)}}_\mu v^\mu_{(0)}\Big(\partial^\alpha U_a+
{2\over3}\sum_{b\not=a}\Big<\partial^\alpha\partial_\mu U_b\Big>_a\cdot 
y^\mu_a\Big)- $$
{}
$$-v^\alpha_{(0)}  v^\lambda_{(0)}\Big(2(\gamma+1) \partial_\lambda U_a+
{2\over3}(\gamma+3)\sum_{b\not=a}\Big<\partial_\lambda\partial_\mu 
U_b\Big>_a\cdot y^\mu_a\Big)+$$
{}
$$+y^\mu_a\Bigg(\gamma\delta^\alpha_\mu{a_{a_0}}_\lambda a^\lambda_{a_0}-
(2\gamma-1)a^\alpha_{a_0}{a_{a_0}}_\mu +2\gamma v^\lambda_{(0)}
\big[\eta_{\lambda\mu}{\dot a}^\alpha_{a_0}-\delta^\alpha_\mu{{\dot a}_{a_0}}
{}_\lambda\big]+$$
{}
$$+\sum_{b\not=a}\Bigg({\partial\over\partial y^0_a}\Big[\gamma \delta^\alpha_\mu
{\partial \over \partial y^0_a}\Big<U_b\Big>_a-
(\gamma+1) \Big<\partial_{(\mu}V^{\alpha)}_b\Big>_a\Big]+$$
{}
$$+2(\gamma+1)v^\lambda_{(0)}\Big[\Big<\partial^\alpha\partial_\mu
 {V_b}_\lambda\Big>_a-\Big<\partial^2_{\lambda\mu}V^{\alpha}_b\Big>_a \Big]+$$
{}
$$+{2\over3}\gamma v^\lambda_{(0)}v^\beta_{(0)}
\Big[\delta^\alpha_\mu\Big<\partial^2_{\lambda\beta}U_b\Big>_a -
\eta_{\lambda\mu}\Big<\partial^\alpha\partial_\beta U_b\Big>_a\Big]\Bigg)\Bigg)+
{\cal O}(|y^\nu_a|^2)+{\cal O}(c^{-6}).\eqno(C3)$$

By expanding all the potentials in Eq.$(C3)$ in power series of 
$ 1/y_{ba_0}$ and   retaining    terms with 
$\sim y^\alpha/|y_{ba_0}|$
only  to the required accuracy, we then obtain:
{}
$$a^\alpha_{(0)}=-\eta^{\alpha\mu}{\partial {\overline U}\over\partial y^\mu_a}+
\delta_a a^\alpha_{(0)}+\delta_{ab} a^\alpha_{(0)}+
\delta_b a^\alpha_{(0)}+\delta_{bc} a^\alpha_{(0)}+{\cal O}(|y^\nu_a|^2)+
{\cal O}(c^{-6}), 
\eqno(C4)$$

\noindent where the post-Newtonian acceleration $\delta_a a^\alpha_{(0)}$ 
due to the gravitational field of the 
body $(a)$ only, may be  given as 
{} 
$$\delta_a a^\alpha_{(0)}=2(\gamma+\beta)U_a \partial^\alpha U_a-
(\gamma+{1\over2})\partial^\alpha\Phi_{1a}+(2\beta-{3\over2})
\partial^\alpha\Phi_{2a}+ $$
{}
$$+(1-\gamma)\partial^\alpha \Phi_{4a}+{1\over2}\partial^\alpha A_a-
(2\gamma+1)v^\alpha_{(0)}
\partial_\mu V^\mu_a +\gamma {v_{(0)}}_\mu v^\mu_{(0)}\partial^\alpha U_a- $$
{}
$$-2(\gamma+1) v^\alpha_{(0)}  v^\mu_{(0)}\partial_\mu U_a-2(\gamma+1) 
{v_{(0)}}_\mu\Big[ \partial^\mu V^\alpha_a-\partial^\alpha V^\mu_a\Big]-$$
{}
$$-2(\gamma+1)\int_ad^3 y'^\nu_a{\hat\rho}_a v'^\alpha_a {v'_a}_\mu
{(y^\mu_a-y'^\mu_a)
\over|y^\nu_a-y'^\nu_a|}-{1\over2}(4\gamma+3) \int_ad^3 y'^\nu_a  
{{\hat\rho}_a\partial^\alpha U_a\over|y^\nu_a-y'^\nu_a|}+$$
{}
$$+{1\over2}  \int_ad^3 y'^\nu_a {\hat\rho}_a\partial_\mu U_a 
{(y^\alpha_a-y'^\alpha_a)(y^\mu_a-y'^\mu_a)\over|y^\nu_a-y'^\nu_a|}+
{\cal O}(c^{-6}).\eqno(C5a)$$

\noindent The term  $\delta_{ab} a^\alpha_{(0)}$ is the 
acceleration due to the interaction of the 
gravitational field of the extended body $(a)$ with the 
external gravitation in the  {\small N}-body system:
{}
$$\delta_{ab} a^\alpha_{(0)}=\sum_{b\not=a}
\Bigg((4\beta-3\gamma-1){m_am_b\over y_{ba_0}}{n^\alpha\over y^2}
+2(\beta-1){m_am_b\over y}{N^\mu_{ba_0}\over {y^2_{ba_0}}}
\big(\delta^ \alpha_\mu+n^\alpha n_\mu\big)+$$
{}
$$+{m_am_b\over y^3_{ba_0}}{\cal P}_{\epsilon\lambda} 
\Big((2\beta+{5\over3}\gamma)\eta^{\alpha\epsilon}n^\lambda+
(\beta-{1\over6})n^\alpha n^\epsilon n^\lambda\Big)+ $$
{}
$$+{m_am_b\over y^4_{ba_0}}\Big((2\beta+\gamma-1)\delta^\alpha_\mu+
2(3\beta+ \gamma-1)N^\alpha_{ba_0}{N_{ba_0}}_\mu\Big) y^\mu_a+ $$ 
{}
$$+3\beta {m_am_b\over y^4_{ba_0}}|y^\nu_a| n^\epsilon n^\lambda
\Big(2\delta^\alpha_\epsilon{N_{ba_0}}_\lambda+ N^\alpha_{ba_0}  
(\eta_{\epsilon\lambda}+5 {N_{ba_0}}_\epsilon
{N_{ba_0}}_\lambda)\Big)\Bigg)+{\cal O}(|y^\nu_a|^2)+ {\cal O}(c^{-6}). 
\eqno(C5b)$$
\noindent Note  that the combination of the post-Newtonian 
parameters in the first term of the expression $(C5b)$   differs 
from that for the well known Nordtvedt effect (\cite{Wil93}, \cite{Nor68b}).
This may provide an independent test for the parameters involved.
The reason  that our third term in this expression   differs from 
the analogous  term derived in \cite{Ash86} is that,
in order to obtain this result Eqs.$(C4$-$C5)$, we  used the consistent  
definitions for the conserved mass density in the proper reference frame. 
Moreover, in  constructing  the Fermi normal coordinates previous authors 
used the incomplete expressions for the spatial coordinate transformations, 
which are  differing from  Eq.$(31a)$. Note that  if one decide to use our 
definitions, the result cited above will take the form of  Eq.$(C5b)$. 
The next term  $\delta_b a^\alpha_{(0)}$ is the
post-Newtonian acceleration  caused by the other 
bodies in the system on the orbit of the body $(a)$
(the effect of the post-Newtonian tidal forces): 
{} 
$$\delta_b a^\alpha_{(0)}=\sum_{b\not=a} y^\mu_a\Bigg(-{3\over2}
{m^2_b\over y^4_{ba_0}}
\Big({1\over3}(14\gamma-4\beta-7)\delta^\alpha_\mu+
3N^\alpha_{ba_0}{N_{ba_0}}_\mu{1\over9}(42\gamma-16\beta-17)\Big)+$$
{}
$$+{m_b\over y^3_{ba_0}}{\cal P}^\alpha_\mu 
\Big[(\gamma+{1\over2}){v_{ba_0}}_\lambda v^\lambda_{ba_0}+
{3\over2}({v_{ba_0}}_\lambda N^\lambda_{ba_0})^2+(4\beta-\gamma-3)E_b\Big]+ $$
{}
$$+\gamma{m_b\over y^3_{ba_0}}{\cal P}_{\epsilon\lambda} 
\Big(\delta^\alpha_\mu v^\epsilon_{ba_0}v^\lambda_{ba_0}-
v^\alpha_{ba_0}v^\epsilon_{ba_0}\delta^\lambda_\mu-
v^\epsilon_{ba_0}{v_{ba_0}}_\mu\eta^{\alpha\lambda}\Big)-$$
{}
$$-{m_b\over y^3_{ba_0}}\Big(\delta^\alpha_\epsilon\eta_{\lambda\mu}-
3N^\alpha_{ba_0}{N_{ba_0}}_\mu{N_{ba_0}}_\epsilon{N_{ba_0}}_\lambda\Big)
v^\epsilon_{ba_0}v^\lambda_{ba_0}+ $$
{}
$$+{m_b\over y^3_{ba_0}}{\cal P}_{\epsilon\lambda} 
\Big({2\over3}\gamma {v_{(0)}}_\beta v^\beta_{(0)} \eta^{\alpha\epsilon}
\delta^\lambda_\mu-
{2\over3}(\gamma+3)v^\alpha_{(0)} v^\epsilon_{(0)}\delta^\lambda_\mu+ $$
{}
$$+{2\over3}\gamma v^\beta_{(0)} v^\lambda_{(0)}
[\delta^\alpha_\mu\delta^\epsilon_\beta-\eta_{\beta\mu}\eta^{\alpha\epsilon}]+
2\gamma v^\beta_{(0)} v^\lambda_{ba_0}
[\delta^\alpha_\mu\delta^\epsilon_\beta-\eta_{\beta\mu}\eta^{\alpha\epsilon}]+$$
{}
$$+2(\gamma+1)v^\beta_{(0)}\delta^\lambda_\mu[{v_{ba_0}}_\beta
\eta^{\alpha\epsilon}-
v^\alpha_{ba_0}\delta^\epsilon_\beta]\Big)\Bigg)+{\cal O}(|y^\nu_a|^2)+
{\cal O}(c^{-6}). \eqno(C5c)$$

\noindent Finally, the last term in the expression $(C4)$ $\delta_{bc} 
a^\alpha_{(0)}$ is the contribution to the equation of motion of the non-linear 
gravitational interaction  of the external bodies with each other
given as follows:
{}
$$\delta_{bc} a^\alpha_{(0)}=-\sum_{b\not=a}\sum_{c\not=a,b}
y^\mu_a\Bigg({m_bm_c\over y^2_{ba_0}y^2_{ca_0}}\Big[3(\gamma-1){\cal P}^\alpha_\mu 
{N_{ba_0}}_\lambda N^\lambda_{ca_0}+$$
{}
$$+(2\beta-1)N^\alpha_{ba_0}{N_{ca_0}}_\mu -N^\alpha_{ca_0}{N_{ba_0}}_\mu\Big]+
{m_bm_c\over y^3_{ba_0}y_{ca_0}}\Big[(2\beta-3\gamma+
{1\over2}){\cal P}^\alpha_\mu +$$
{}
$$+{1\over2}\gamma{\cal P}_{\epsilon\lambda} 
\Big(\eta^{\alpha\epsilon}{N_{ca_0}}_\mu N^\lambda_{ca_0}+\delta^\epsilon_\mu
N^\alpha_{ca_0} N^\lambda_{ca_0}-(\delta^\alpha_\mu+ N^\alpha_{ca_0} 
{N_{ca_0}}_\mu) N^\epsilon_{ca_0} N^\lambda_{ca_0}\Big)\Big]+$$
{}
$$+{m_bm_c\over y^2_{ba_0}y^2_{cb_0}}\Big[{1\over2}
{\cal P}^\alpha_\mu {N_{ba_0}}_\lambda N^\lambda_{cb_0}+
\gamma\delta^\alpha_\mu {N_{ba_0}}_\lambda N^\lambda_{cb_0}-$$
{}
$$-(\gamma+{1\over2})(N^\alpha_{ba_0} {N_{cb_0}}_\mu+{N_{ba_0}}_\mu 
N^\alpha_{cb_0})\Big]+(2\beta-{3\over2}){m_bm_c\over y^3_{ba_0}y_{cb_0}}
{\cal P}^\alpha_\mu\Bigg)+{\cal O}(|y^\nu_a|^2)+
{\cal O}(c^{-6}).\eqno(C5d)$$

Thus, the equations presented in this Appendix are represent the motion of a 
test body in the Fermi-normal-like coordinates chosen in the proper reference 
frame of a body $(a)$. Together with the coordinate transformations Eq.(22) 
this is the general solution of the gravitational  {\small N}-body problem.
 
\vskip 18pt\noindent 
{\large\bf Appendix D:    Astrophysical Parameters Used in the Paper.}
\vskip 11pt\noindent
In this Appendix we  present the astrophysical parameters used in the 
calculations of the gravitational effects for the {\it Mercury Orbiter} mission 
in Section {\small V}:
$$\hbox{Solar radius}:  R_\odot  = 695,980 \hbox{ km},$$  
$$\hbox{Solar gravitational constant}:  \mu_\odot  = c^{-2}G M_\odot=1.4766 
\hbox{ km},$$  
$$\hbox{Solar quadrupole coefficient (Brown {\it et al.}, 1989)}:  
J_{2\odot} = (1.7\pm0.17)\times 10^{-7},$$ 
$$\hbox{Solar rotation period}:  \tau_\odot =  25.36 \hbox{ days},$$ 
$$\hbox{Mercury's mean distance}:  a_M  = 0.3870984  \hbox{ AU}= 
57.91\times 10^6 \hbox{ km},$$ 
$$\hbox{Mercury's radius}: R_M  = 2,439  \hbox{ km},$$ 
$$\hbox{Mercury's gravitational constant}:  \mu_M = c^{-2}G M_M= 
1.695\times 10^{-7}\mu_\odot,$$
$$\hbox{Mercury's sidereal period}: T_M  = 0.241  \hbox{ yr} =
87.96 \hbox{ days},$$ 
$$\hbox{Mercury's rotational period}: \tau_M  =  59.7 \hbox{ days },$$ 
$$\hbox{Eccentricity of Mercury's orbit}: e_M  =  0.20561421,$$ 
$$\hbox{Jupiter's gravitational constant}:  \mu_J = 
9.547\times 10^{-4}\mu_\odot,$$ 
$$\hbox{Jupiter's sidereal period}: T_J  = 11.865  \hbox{ yr},$$ 
$$\hbox{Astronomical Unit}: AU = 1.49597892(1)\times 10^{13}  \hbox{ cm}.$$

\end{document}